\documentclass[twocolumn,prc,showpacs,preprintnumbers,
               unsortedaddress,amsmath,amssymb,floatfix]{revtex4}
\usepackage{graphicx}
\usepackage{dcolumn}
\usepackage{bm}
\usepackage{longtable}

\newcommand{\beq}{\begin{equation}}
\newcommand{\eeq}{\end{equation}}
\newcommand{\beqn}{\begin{eqnarray}}
\newcommand{\eeqn}{\end{eqnarray}}
\newcommand{\btab}{\begin{tabular}}
\newcommand{\etab}{\end{tabular}}
\newcommand{\etal}{{\em{et al.}}}

\usepackage{multirow,amsmath,booktabs}

\begin{document}

\title{A Second Relativistic Mean Field and Virial Equation of State for Astrophysical Simulations}

\author{G.~Shen\footnote{e-mail:  gshen@lanl.gov}}
\affiliation{Theoretical Division, Los Alamos National Lab, Los Alamos, NM 87545, USA}
\affiliation{Center for the Exploration of Energy and Matter and Department of Physics,
Indiana University Bloomington, IN 47405, USA}
\author{C.~J.~Horowitz\footnote{e-mail:
horowit@indiana.edu} } \affiliation{Center for the Exploration of Energy and Matter and
Department of Physics, Indiana University Bloomington, IN 47405, USA}
\author{E. O'Connor\footnote{e-mail:
evanoc@tapir.caltech.edu}}
\affiliation{TAPIR, Mail Code 350-17, California Institute of Technology, Pasadena, CA, 91125, USA}

\date{\today}
\begin{abstract}
We generate a second equation of state (EOS) of nuclear
matter for a wide range of temperatures, densities, and proton
fractions for use in supernovae, neutron star mergers, and black hole formation simulations.  We employ full relativistic mean field (RMF)
calculations for matter at intermediate density and high density,
and the Virial expansion of a non-ideal gas for matter at low density.  For this EOS we use the RMF effective interaction FSUGold, whereas our earlier EOS was based on the RMF effective interaction NL3.   The FSUGold interaction has a lower pressure at high densities compared to the NL3 interaction.  We calculate the resulting EOS at over 100,000 grid points in the temperature range $T$ = 0 to 80 MeV, the density range $n_B$ = 10$^{-8}$ to 1.6 fm$^{-3}$, and the proton fraction range $Y_p$ = 0 to 0.56.  We then interpolate these data points using a suitable scheme to generate a thermodynamically consistent equation of state table on a finer grid.  We discuss differences between this EOS, our NL3 based EOS, and previous EOS by Lattimer-Swesty and H. Shen \etal\ for the thermodynamic properties, composition, and neutron star structure.  The original FSUGold interaction produces an EOS, that we call FSU1.7, that has a maximum neutron star mass of 1.7 solar masses.  A modification in the high density EOS is introduced to increase the maximum neutron star mass to 2.1 solar masses and results in a slightly different EOS that we call FSU2.1. The EOS tables for FSU1.7 and FSU2.1 are available for download.

\end{abstract}


\maketitle

\section{introduction}

The evolution of compact stellar objects provides a unique laboratory to study matter at extremely high densities, temperatures, and/or neutron fractions.  Neutron star (NS) binary systems emitting X-ray bursts or super bursts provide a wealth of information on the NS crust, its composition, and transport properties \cite{crust}.  Core collapse supernovae (SN) release gigantic amounts of gravitational energy through the emission of neutrinos.  Their dynamics depends on detailed understanding of neutrino interactions in dense matter \cite{sn}.  NS-NS and black hole-NS mergers produce strong gravitational waves that depend on the equation of state (EOS) of dense matter \cite{merger}.   The EOS gives the pressure as a function of density, temperature and proton fraction.  The historic detection of gravitational waves is anticipated soon, as the Advanced LIGO and VIRGO detectors become operational \cite{LIGO}. 

For the past two decades, there have been several attempts to construct complete EOS tables over wide range of temperatures, densities, and compositions. However most realistic supernova simulations have used only two EOS tables.  The Lattimer-Swesty (L-S) equation of state \cite{LS}, that is based on a compressible liquid drop model with a Skyrme force, and the H. Shen, Toki, Oyamatsu and Sumiyoshi (S-S) equation of state \cite{Shen98a,Shen98}, that is based on a relativistic mean field (RMF) model with Thomas-Fermi approximation and variational method.  Recently, Hempel and Schaffner-Bielich have built an EOS table by matching a nuclear statistical equilibrium model, at low densities, directly to uniform nuclear matter at high densities \cite{hempel}.  In addition, the S-S EOS has been extended to include hyperons \cite{hyperons1,hyperons2}.

In a series of papers \cite{paper0,paper1,paper2,paper3},  two of the present authors presented a new complete EOS.  This is based on an RMF model, NL3 \cite{NL3}, to self-consistently calculate nonuniform matter at intermediate density and uniform matter at high density.  Matter at low density is described with a Virial expansion for a non-ideal gas of nucleons \cite{virialneut} and nuclei \cite{virialnuc}.  This Virial approach uses elastic scattering phase shifts and nuclear masses as input and includes Coulomb corrections that can be important for neutrino interactions \cite{ioncorrelations,ioncorrelationsB}.  The Virial expansion is exact in the low density limit.  Altogether our RMF and Virial EOS models cover the large range of temperatures, densities, and proton fractions  necessary for astrophysical simulations. This EOS table is available for download \cite{paper3}.

In this work, we use a different RMF effective interaction FSUGold \cite{FSUGold} to study matter at intermediate and high densities. This replaces the NL3 effective interaction used in our previous work. The low density EOS from the Virial expansion is unchanged. FSUGold reproduces ground state properties of nuclei across the periodical table very well. The differences between NL3 and FSUGold have been extensively studied for example, in \cite{FSUNL3}. It was observed that FSUGold is consistent with most known constraints, which include the universal behavior of dilute Fermi gases with large scattering lengths, heavy-ion experiments that probe both the low- and high-density domain of the equation of state. NL3 may not satisfy these constraints as good as FSUGold. However, there are still large controversies in these theoretical and experimental constraints. Therefore it is very useful to have EOS produced with both effective interactions. The astrophysical simulations with the two EOS could potentially provide more constraints on the EOS, which are correlated with various astrophysical observables. The FSUGold EOS is considerably softer than NL3 at high density, but stiffer at subnuclear density. The original FSUGold EOS supports a maximum NS mass of 1.7 $M_\odot$, which is in contradiction to the recent observation of 2 solar mass NS \cite{2solar}. We introduce a modification term to the original FSUGold interaction that increases the pressure at high densities so that the maximum NS mass is increased to 2.1 $M_\odot$.

The paper is organized as follows. In section 2 we present the RMF models for NL3 and FSUGold effective interactions.  Then a modification to the FSUGold effective interaction is presented, which results in a maximum NS mass of 2.1 $M_\odot$. We also summarize numerical details for the interpolation scheme we use to generate the full EOS table. Section 3 shows results for our EOS, including various thermodynamic properties and the composition. The $T = 0$ beta equilibrium EOS is also presented and used to calculate the NS mass-radius relation. We compare our EOS and existing L-S and S-S EOS tables. Section 4 presents a summary of our results and gives an outlook for future work.  In the appendix the public access to our EOS table is given.

\section{Formalism}
In this section we describe our relativistic mean field interaction, computational method, and interpolation scheme.
 
\subsection{Relativistic mean field interactions: FSUGold and NL3}

In this subsection we discuss the RMF interactions FSUGold and NL3.  At low density the Virial expansion is the same in both EOS and has been discussed in a previous work \cite{paper2}. We note there are more low temperature ($<$ 1 MeV) data in Virial gas in the FSU EOS, which only induces small difference between NL3 and FSU EOS. The comparison between NL3 and FSUGold effective interactions has been discussed extensively in Ref.~\cite{IUFSU}.  For completeness, we briefly present the RMF models to motivate this work.

The RMF theory that we use is based on 
Ref.~\cite{Mueller:1996pm} supplemented with an isoscalar-isovector coupling as introduced in Ref. \cite{Horowitz:2000xj}. The basic ansatz of the RMF theory is a Lagrangian density where nucleons interact via the exchange of sigma- ($\phi$), omega-
($V_\mu$), and rho- (${\bf b}_\mu$) mesons, and also photons
($A_\mu$), given by~\cite{Horowitz:2000xj,Mueller:1996pm}
\begin{eqnarray}
{\cal L}_{\rm int} &=&
\bar\psi \left[g_{\rm s}\phi   \!-\! 
         \left(g_{\rm v}V_\mu  \!+\!
    \frac{g_{\rho}}{2}{\mbox{\boldmath $\tau$}}\cdot{\bf b}_{\mu} 
                               \!+\!    
    \frac{e}{2}(1\!+\!\tau_{3})A_{\mu}\right)\gamma^{\mu}
         \right]\psi \nonumber \\
                   &-& 
    \frac{\kappa}{3!} (g_{\rm s}\phi)^3 \!-\!
    \frac{\lambda}{4!}(g_{\rm s}\phi)^4 \!+\!
    \frac{\zeta}{4!}   g_{\rm v}^4(V_{\mu}V^\mu)^2 \nonumber \\
    &+&
   \Lambda_{\rm v} g_{\rho}^{2}\,{\bf b}_{\mu}\cdot{\bf b}^{\mu}
           g_{\rm v}^{2}V_{\nu}V^\nu\;.
 \label{LDensity}
\end{eqnarray} The values for various parameters in the interacting lagrangian can be found in Table I of Ref.~\cite{IUFSU}. Here we duplicate the bulk properties of infinite nuclear matter in Table~\ref{Table3}. The incompressibility, symmetry energy and its slope at nuclear saturation density predicted by FSUGold are considerably smaller than NL3, {\it i.e.}, the FSUGold EOS is much softer than NL3 at saturation density and above. Therefore FSUGold gives a smaller maximum NS mass, which will be discussed in following sections. However, we note that FSUGold gives slightly stiffer nuclear EOS at subnuclear density than NL3. As a result, FSUGold predicts a smaller neutron skin thickness in $^{208}$Pb, 0.21 fm compared to 0.28 fm given by NL3 \cite{IUFSU}. The experimental constraints on these nuclear properties at best can only be inferred indirectly. The incompressibility is constrained by giant resonance of nuclei, while the symmetry energy and its derivative can be constrained for example by studying the neutron skin thickness of $^{208}$Pb, or comparing with microscopic calculations using chiral perturbation theory or quantum Monte Carlo method. Both constraints on FSUGold and NL3 have been studied extensively in Ref.~\cite{IUFSU}. Again, FSUGold is consistent with these constraints.

To better describe low density neutron matter, which is close to a unitary gas, we also introduced a density dependent coupling $g_{\rm s}$ between scalar mesons and nucleons for FSUGold, which is the same as that for NL3 in Ref. ~\cite{paper1}. Then we solve the Hartree problem at finite temperature for non-uniform nuclear matter at intermediate densities, and uniform nuclear matter at high densities, following Ref.~\cite{paper1}.

\begin{table}
\begin{tabular}{|l||c|c|c|c|c|}
 \hline
 Model & $\rho_{0}~({\rm fm}^{-3}) $ & $\varepsilon_{0}$~(MeV) 
           & $K_{0}$~(MeV) & $J$~(MeV) & $L$~(MeV) \\
\hline
 \hline
 NL3    &  0.148  & $-$16.24 & 271.5 & 37.29 & 118.2 \\ 
 FSU     &  0.148  & $-$16.30 & 230.0 & 32.59 & 60.5  \\

\hline
\end{tabular}
\caption{Bulk parameters of infinite nuclear matter at saturation density 
              $\rho_{_{0}}$. The quantities $\varepsilon_{_{0}}$ 
              and $K_{0}$ represent the binding energy per 
              nucleon and incompressibility coefficient of 
              symmetric nuclear matter, whereas $J$ and 
              $L$ represent the symmetry energy and derivative of the symmetry energy with respect to density at $\rho_{_0}$.}
\label{Table3}
\end{table}

\subsection{\label{modifiedFSU}Modified FSUGold effective interaction: FSU2.1}

The recent observation of a $1.97\pm 0.04$ $M_\odot$ NS \cite{2solar} puts a new constraint on the maximum mass allowed by effective interactions.  The original FSUGold interaction predicts a maximum NS mass of only 1.7 $M_\odot$. The FSUgold interaction is fitted to many observables near nuclear density, and the extrapolation to high densities is incompletely constrained.  We consider a modification at high densities that increases the maximum NS mass.  We assume additional repulsion between nucleons at short distances that is primarily isoscalar.  This will lead to a correction to the pressure that is approximately independent of proton fraction.  Furthermore, matter is degenerate at high densities so that the correction should be nearly independent of temperature.  For simplicity we add the following term to the pressure,

\beq\label{dp} dP\ = a (\rho^2- {\hat\rho_0}^2), \eeq
where $\rho$ is the baryon density and $\hat\rho_0$ is a constant corresponding to 0.2 fm$^{-3}$.  Note that the extra pressure is zero for $\rho < \hat\rho_0$. The corresponding change to the energy is determined from the density integral of $dP$. The constant $a$=$2\times 10^{-5}$ MeV$^{-2}$ was chosen to increase the maximum NS mass to about 2.1 M$_\odot$.  We refer to the EOS built with the original FSUGold effective interaction as FSU1.7, while the EOS with the modified effective interaction will be called FSU2.1. 

This procedure makes no corrections at low densities.  This leaves unchanged all of the FSUgold predictions for nuclear observables.  An alternative, and more consistent, procedure is to refit the effective interaction with a smaller $\zeta$ coupling in Eq. (1).  This will increase the pressure at high densities and lead to a larger maximum NS mass, \cite{IUFSU}.  An equation of state with this modified interaction will be presented in future work.  However, we expect our simple prescription to give similar results.

\subsection{Computational methods}

We calculate the FSU EOS for many points in the parameter spaces of density $n_B$, temperature $T$, and proton fraction $Y_p$ using the procedure in our previous paper \cite{paper1}. There are a total of 107,000 points in the 3-dimensional parameter spaces of  $T$, $Y_p$, and $n_B$.  The overall FSU EOS took about 200,000 CPU hours  and was run primarily on the Teragrid supercomputer cluster Ranger.  Details about these parallel computations have been discussed in previous work \cite{paper1,paper2}.

\subsection{\label{smooth}Bicubic interpolation}

As discussed in our previous paper \cite{paper3}, we use a hybrid method of interpolation to generate an EOS table on a finer grid of density and temperature points.  The thermodynamic pressure $P_{th}$ can be obtained numerically from the free energy per baryon $F/A$, 
\beq
\label{fpressure} P_{th}\ =\ n_b^2
\left(\frac{\partial (F/A)}{\partial n_b}\right)_{T,Y_p}. 
\eeq
For the zero temperature EOS, we smooth the pressure as a function of density for the calculated data points in Table \ref{tab:1}. Note the upper limit in densty is 10$^{0.2}$ and 10$^{0.4}$ fm$^{-3}$ for FSU2.1 and FSU1.7, respectively. This smoothing involves comparing the pressure at a given density to the pressure interpolated from neighboring points. We smooth the pressure in a 10 points per decade table by removing points that differ significantly ($>10^{-3}$) from the geometric mean of neighboring points and replace them with interpolated one.  Then we interpolate these smoothed pressures to create a finer table with 40 points per decade of density. The resulting energy at zero temperature is easily obtained by integration of these pressures with respect to density. The pasta phase has very close thermodynamic properties as normal nuclei \cite{paper0}. Therefore we do not expect the smoothing procedure will lose important features of pasta phase. The purpose of smoothing is just to smooth the fluctuations in numerical evaluation of pressure and entropy.

\begin{table}[h]
\centering \caption{Range of temperature $T$, baryon density $n_B$ and proton
fraction $Y_p$ in the coarse EOS table.} \label{tab:1}\btab{cccc}
\hline \hline
Parameter & minimum & maximum & number of points \\
 \hline
$T$ [MeV] & 0, 10$^{-0.8}$ & 10$^{1.9}$ & 36 \\

log$_{10}$(n$_B$) [fm$^{-3}$] &-8.0 & 0.2 (0.4) & 83 (85) \\

Y$_P$  & 0, 0.05  & 0.56 & 53 \\

\hline \etab
\end{table}

Next, we calculate the entropy at finite temperatures $T<12.5$ MeV.
From the free energy at temperature $T$, density $n_B$, and proton fraction $Y_p$ as in Table~\ref{tab:1}, the entropy per baryon $s_{th}$ can be obtained numerically,
\beq\label{fentropy} s_{th}\ \equiv\ S/A\ =\ -\left(\frac {\partial (F/A)}{\partial
T}\right)_{n_b, Y_p}. \eeq 
Finally, the energy per baryon $e_{th}$ is
\beq
\label{fenergy} e_{th} = F/A - T s_{th}.
\eeq
For the finite temperature EOS with $T<12.5$ MeV, we smooth the entropy, as a function of density, while ensuring thermodynamic stability. Then we perform monotonic cubic Hermite interpolation \cite{COtt} on this small table of entropy with ten points per decade, to generate a larger table with 40 points per decade in both the temperature and density axes, as in Table \ref{tab:phasespace2}. Now we integrate this smoothed entropy table as a function of temperature to get values of the free energy (adding the energy at zero temperature to make the full free energy). Thus we obtain the free energy on this finer 40 by 40 points per decade grid in a thermodynamically consistent fashion.

Note that for higher temperatures $T \geq$ 12.5 MeV, matter is uniform for all proton fractions and densities.  For uniform matter, all thermodynamic quantities can be obtained directly from RMF calculations, with good thermodynamic consistency.

\begin{table}[h]
\centering \caption{Range of temperature $T$, density $n_B$, and proton
fraction $Y_p$ in the finely spaced interpolated EOS table.} \label{tab:phasespace2}\btab{cccc}
\hline \hline
Parameter & minimum & maximum & number of points \\
 \hline
T [MeV] & 0, 10$^{-0.8}$ & 10$^{1.875}$ & 109 \\

log$_{10}$(n$_B$) [fm$^{-3}$] &-8.0 & 0.175 (0.375) & 328 (336) \\

Y$_P$  & 0, 0.05  & 0.56 & 1(Y$_P$=0)+52 \\

\hline \etab
\end{table}

Finally, we carry out bicubic interpolation of the previous free energy values (as in Table \ref{tab:phasespace2}) to generate the entropy and pressure by thermodynamic derivatives Eqs.~(\ref{fpressure},\ref{fentropy}). This prescription guarantees the monotonicity of entropy and pressure in the final table, and conserves the first law of thermodynamics in adiabatic compression tests.

We first apply bicubic interpolation \cite{numericalrecipe} for the free energy.  
The first derivatives on the grid points are generated from monotonic cubic Hermite interpolation \cite{COtt}. The second derivative - the cross derivative $\partial^2 F/\partial n \partial T$, on the grid points is generated as in Ref.~\cite{numericalrecipe}. The bicubic interpolation can then fit free energies with cubic functions in temperature and density coordinates, and provides the first and second (cross) derivatives. Then the entropy and pressure are obtained from Eqs.~(\ref{fpressure},\ref{fentropy}). Finally, the boundary points at the highest temperature or density are discarded to avoid boundary artifacts in the interpolation.  We note that our interpolation procedure is general and should work independent of the approximations used to calculate the original free energy values.  We test the thermodynamic consistency of our table in Sec. \ref{sec.test.results}.

The neutron chemical potential is derived from $\mu_n=\partial(n_B F/A)/\partial n_n|_{T,n_p}$, while $\mu_p = (F/A +p/n_B - \mu_n (1-Y_p))/Y_p$.

\section{Results}
In this section we present results for the free energy, composition, pressure, NS structure, and adiabatic index predicted by our FSU EOS.  We compare our results for the FSU EOS together with NL3 EOS to those for the Lattimer-Swesty or H. Shen \etal\  EOS.

 \subsection{Free energy and Composition}

In this subsection we discuss the free energy, average mass number of heavy nuclei, and mass fractions of different components ($n$, $p$, $\alpha$ and nuclei) for the FSU EOS. 
In Fig.~\ref{fig:freeenergy}, the free energy per baryon $F/A$ for matter at
$T$ = 1, 3.16, 6.31, and 10 MeV with different proton fractions is
shown as a function of density $n_B$.  The free energy $F/A$ is obtained from
Virial gas, nonuniform Hartree mean field, and
uniform matter calculations. This figure resembles Figure 1 in our previous paper for the NL3 EOS \cite{paper3}. The transitions between the Virial expansion and the Hartree calculations (vertical dashed lines), or between the Hartree calculations and the uniform matter calculations (vertical solid lines) are at the densities where Hartree or uniform matter calculations give the lower free energy. The transition between Virial EOS and RMF EOS is pretty smooth, as discussed in Ref.~\cite{paper2}. However, there are errors in calculating the pressure and entropy numerically. It is the purpose of smoothing procedure outlined in Sec.~\ref{smooth} to remove these numerical errors.

Figure~\ref{fig:massnumber} shows the average mass number $\bar{A}$ of
heavy nuclei (with $A>4$) versus baryon density at different temperatures and
proton fractions.  Note that $\alpha$ particles are not counted as heavy nuclei.  Let us first look at the upper left panel (a), where $T$ = 1 MeV.  Nuclear shell effects give rise to several approximate plateaus in $\bar{A}$ vs density for each $Y_p$, for example $\bar{A}$ $\sim$ 50, $\sim$ 80 and $\sim$ 110.   Usually $\bar{A}$ is larger in matter with smaller $Y_p$. There are oscillations in $\bar{A}$ in the Hartree mean field regime.  This is due to both nuclear shell effects and small errors in the free energy minimization due to our using a finite step in the Wigner Seitz cell size \cite{paper1}.  The average mass $\bar{A}$ can be as large as 4,000 at high density.   These large $\bar A$ values represent shell states or spherical pasta configurations \cite{paper0}.  At higher temperatures, as shown in the other panels (b-d) of Fig. \ref{fig:massnumber}, $\bar{A}$ grows rapidly to several thousand in a narrow range of density before $\bar{A}$ drops abruptly at the transition to uniform matter.

\begin{widetext}

\begin{figure}[htbp]
 \centering
 \includegraphics[height=8.5cm,angle=-90]{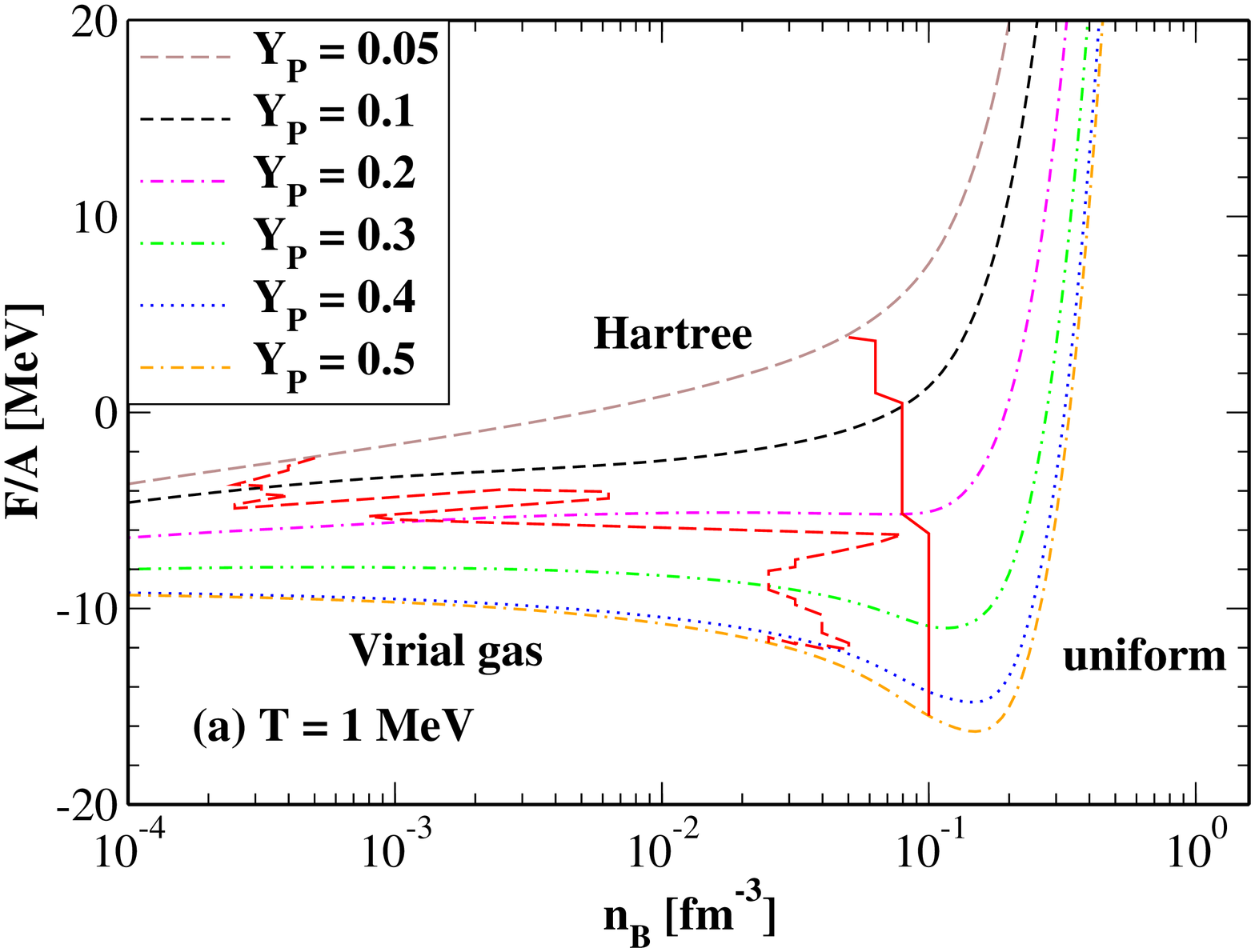}
 \includegraphics[height=8.5cm,angle=-90]{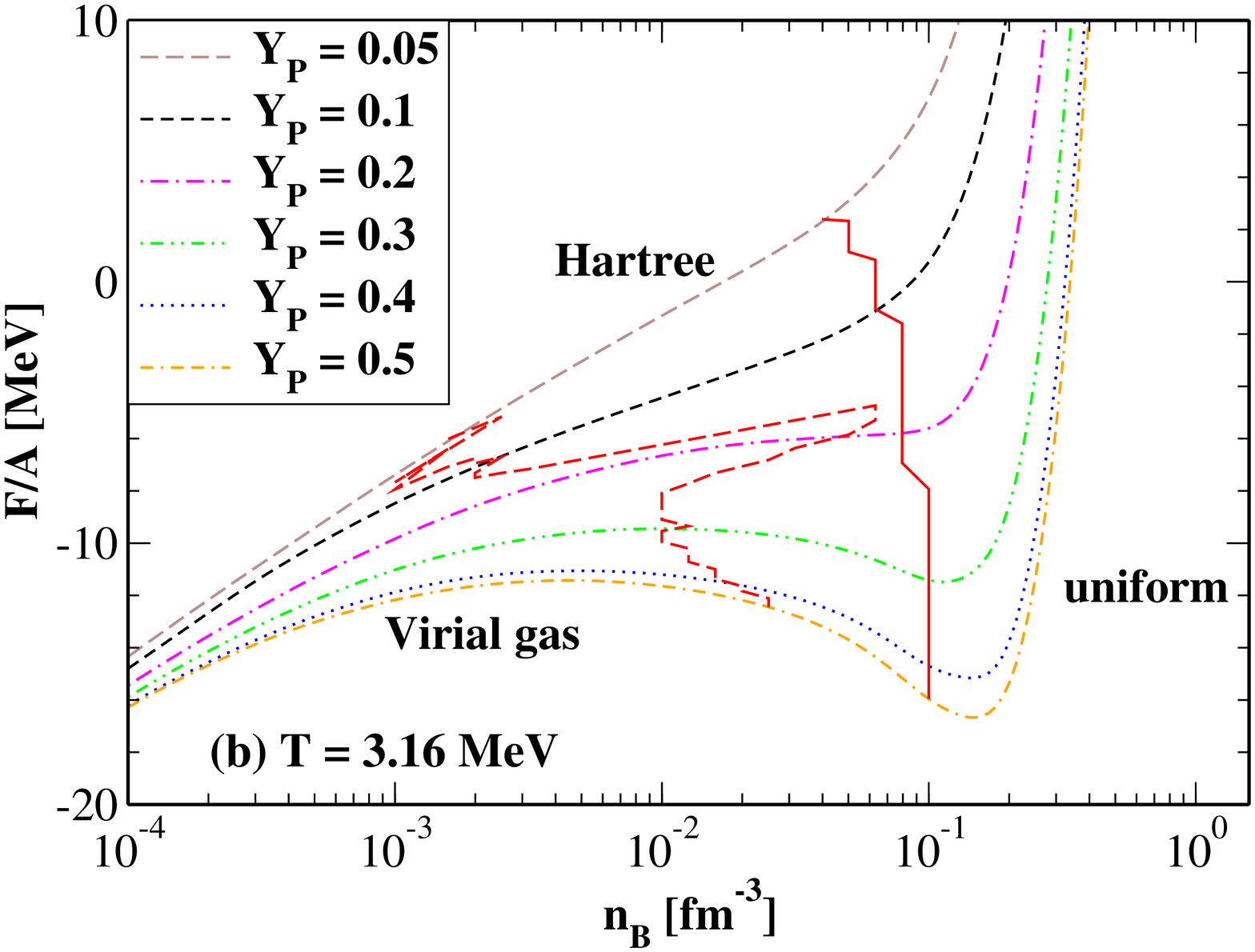}
 \includegraphics[height=8.5cm,angle=-90]{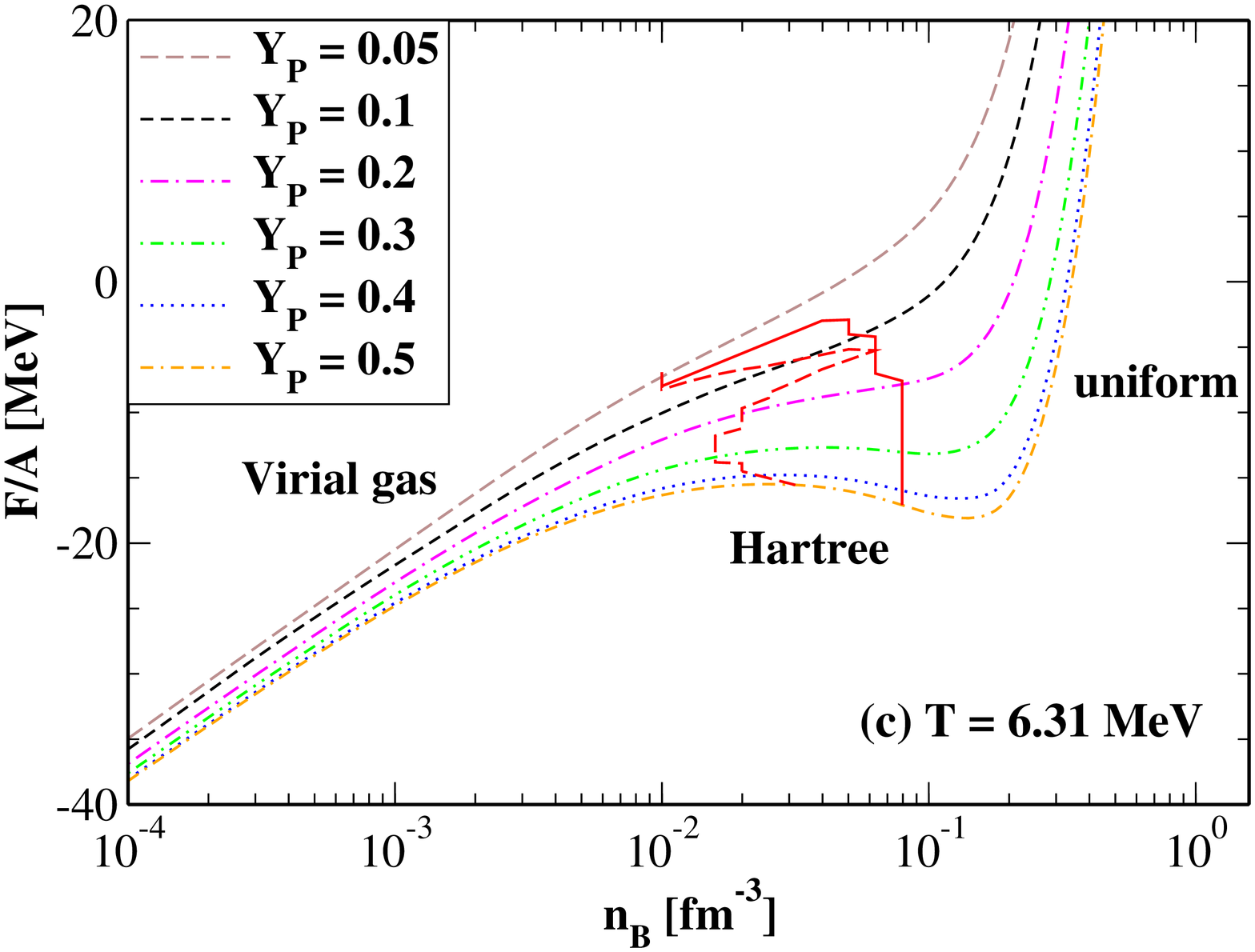}
 \includegraphics[height=8.5cm,angle=-90]{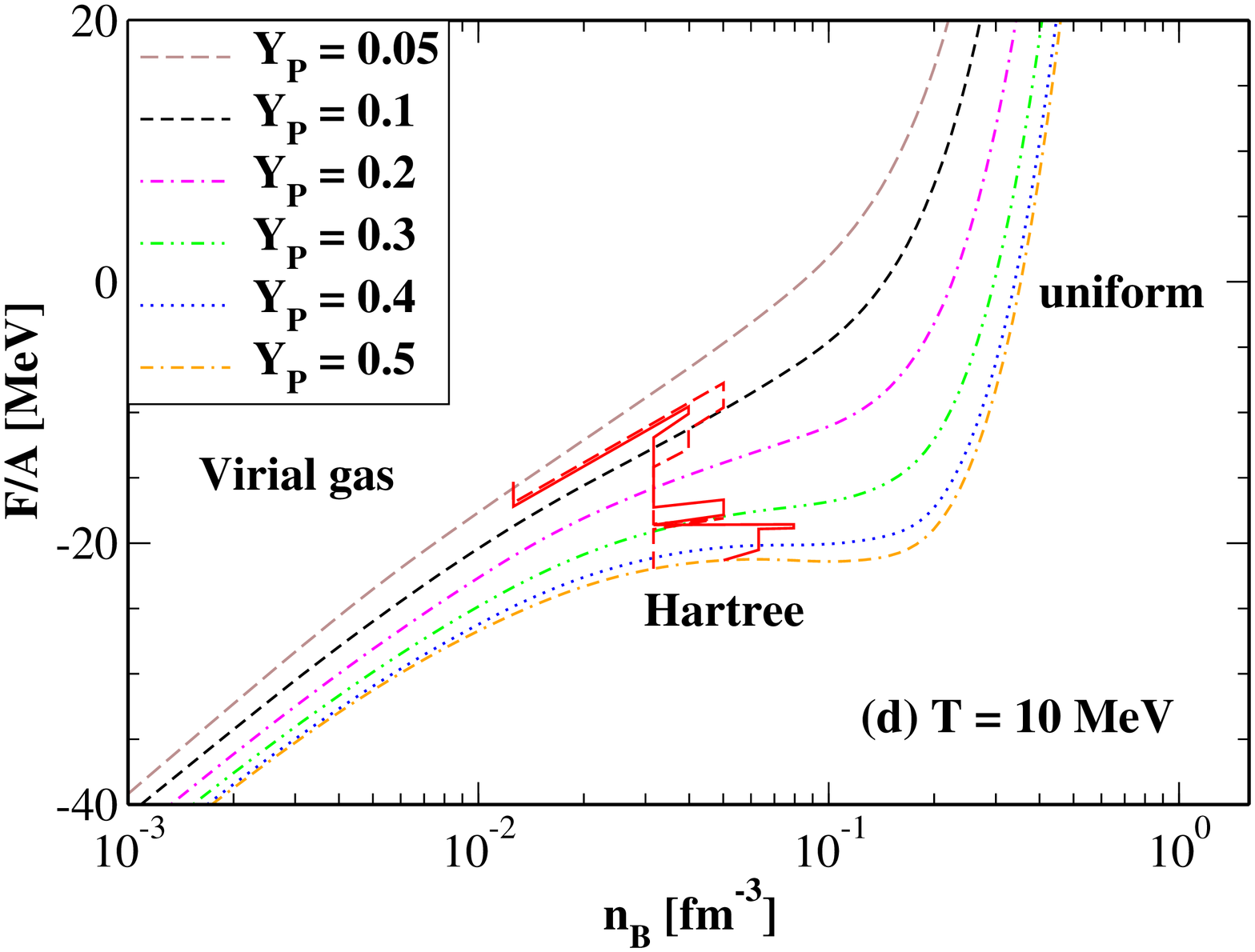}

\caption{(Color online) Free energy per baryon at temperatures of $T$ = 1 (a), 3.16 (b), 6.31 (c), and 10 (d) MeV. The proton fraction ranges from $Y_p$ = 0.05 to 0.5. The transitions between the Virial expansion and the Hartree calculations are vertical dashed lines.  The transitions between Hartree calculations and the uniform matter calculations are vertical solid lines. }\label{fig:freeenergy}
\end{figure}


\begin{figure}[htbp]
 \centering
 \includegraphics[height=8.5cm,angle=-90]{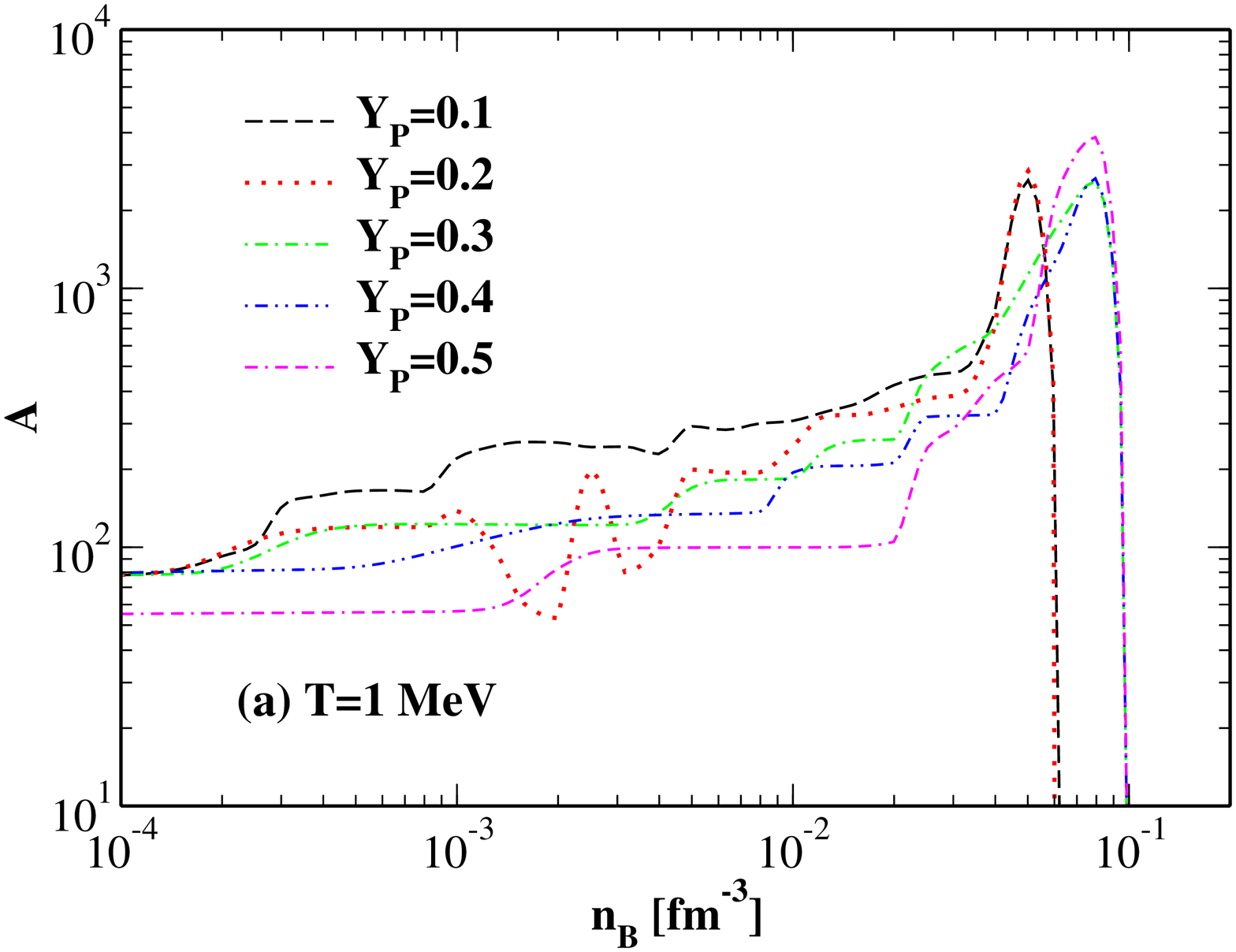}
 \includegraphics[height=8.5cm,angle=-90]{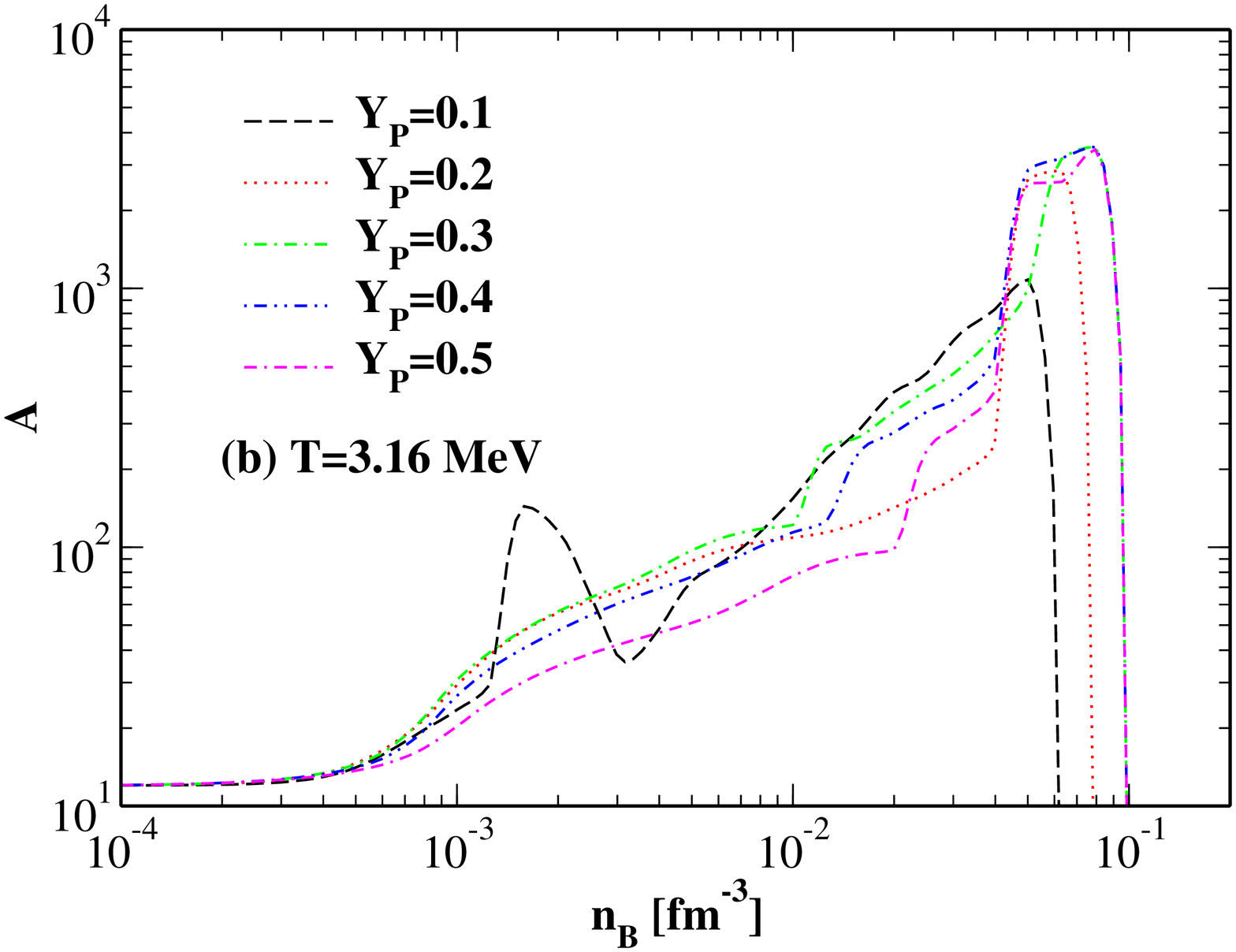}
 \includegraphics[height=8.5cm,angle=-90]{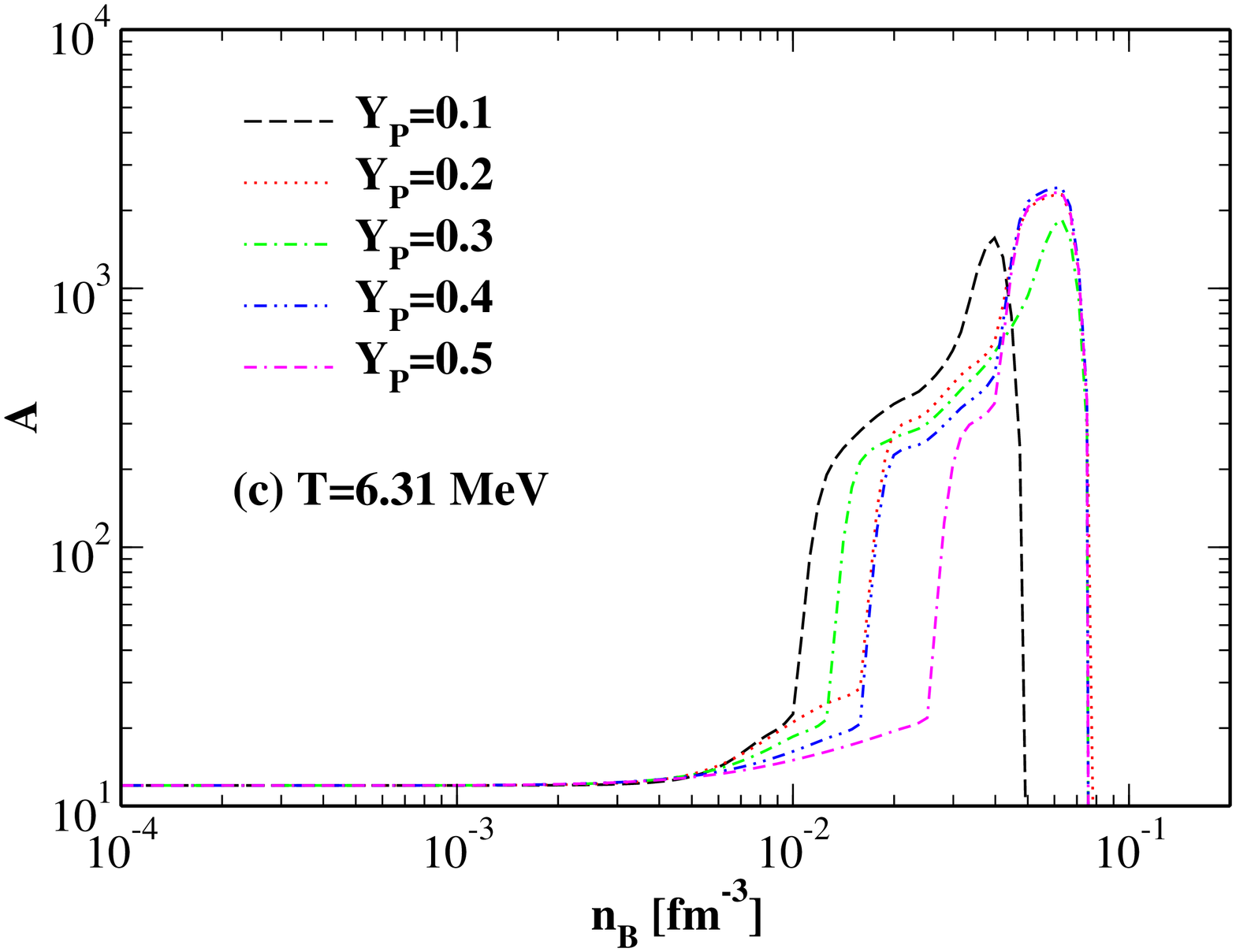}
 \includegraphics[height=8.5cm,angle=-90]{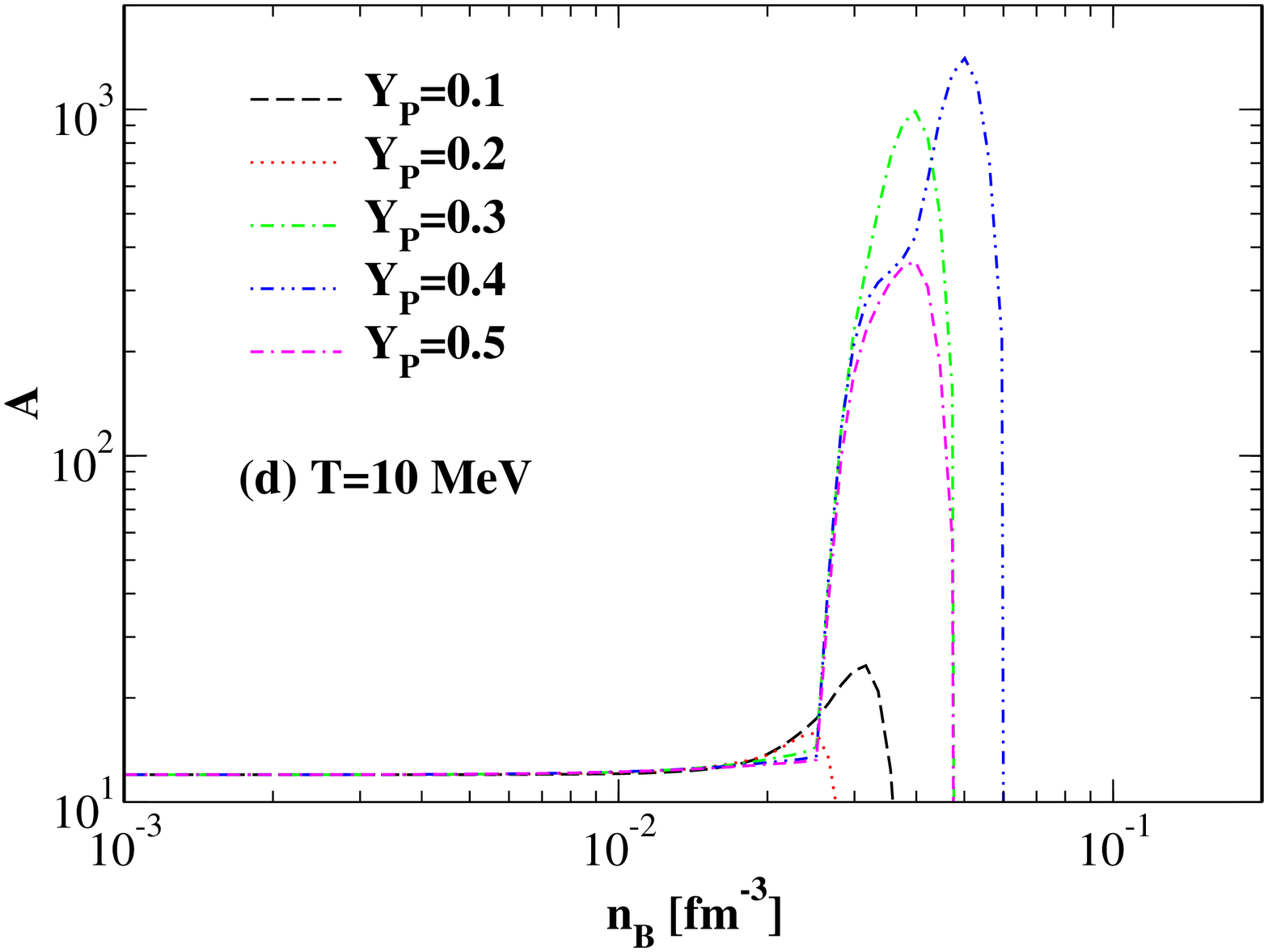}

\caption{(Color online) Average mass number of heavy nuclei at temperatures of $T$ = 1 (a), 3.16 (b), 6.31 (c), and 10 (d) MeV. The proton fraction ranges from $Y_p$ = 0.1 to 0.5.}\label{fig:massnumber}
\end{figure}


\begin{figure}[htbp]
 \centering
 \includegraphics[height=13cm,angle=-90]{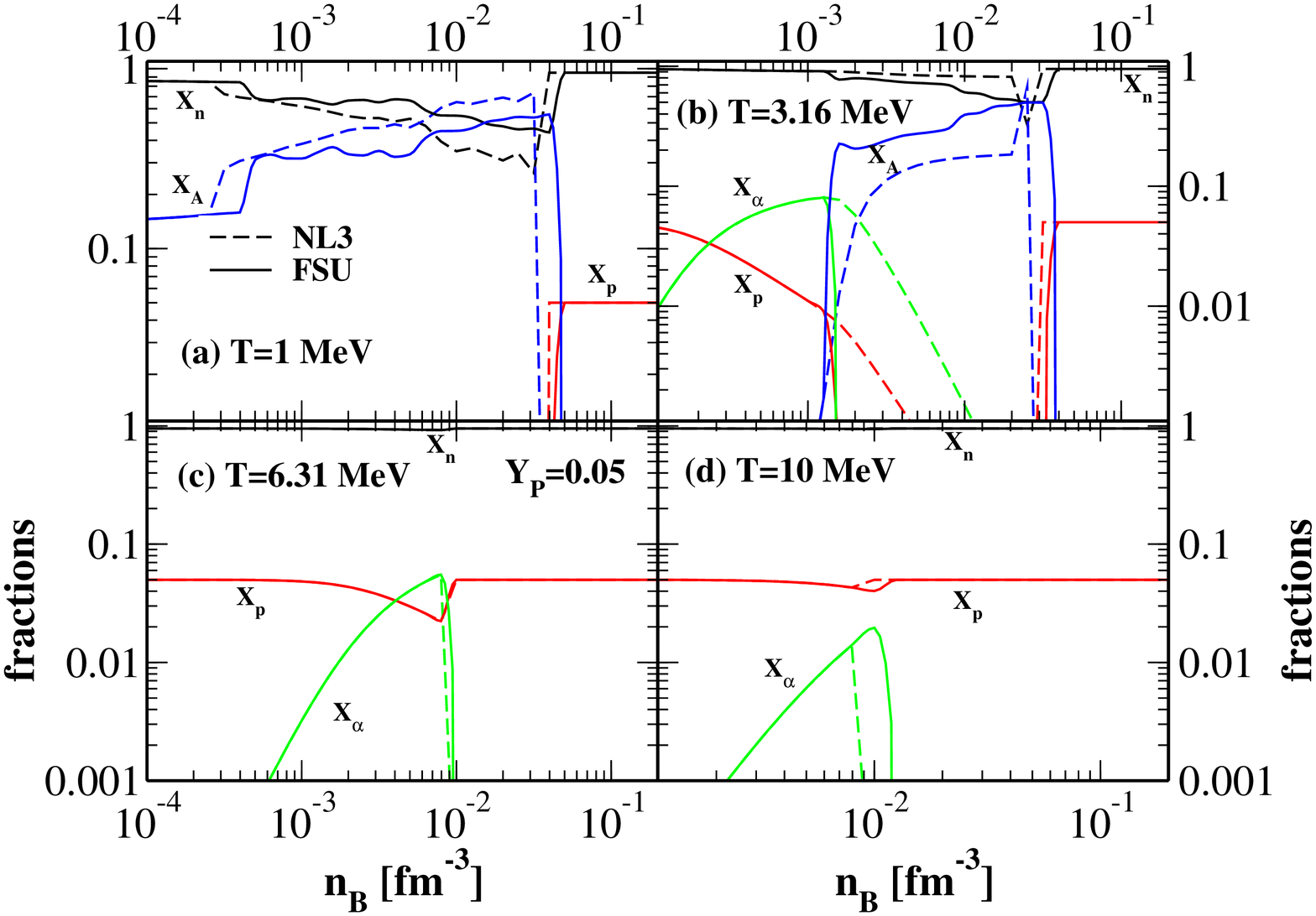}

\caption{(Color online) Mass fractions of unbound neutrons, unbound protons, $\alpha$
particles and heavy nuclei versus baryon density at four different
temperatures $T$ = 1 (a), 3.16 (b), 6.31 (c), and 10 (d) MeV, and fixed proton fraction $Y_p$ =
0.05. The solid lines are for FSU(1.7) EOS while the dashed lines are for NL3 EOS.}\label{fig:fractiony05}
\end{figure}

\begin{figure}[htbp]
 \centering
 \includegraphics[height=13cm,angle=-90]{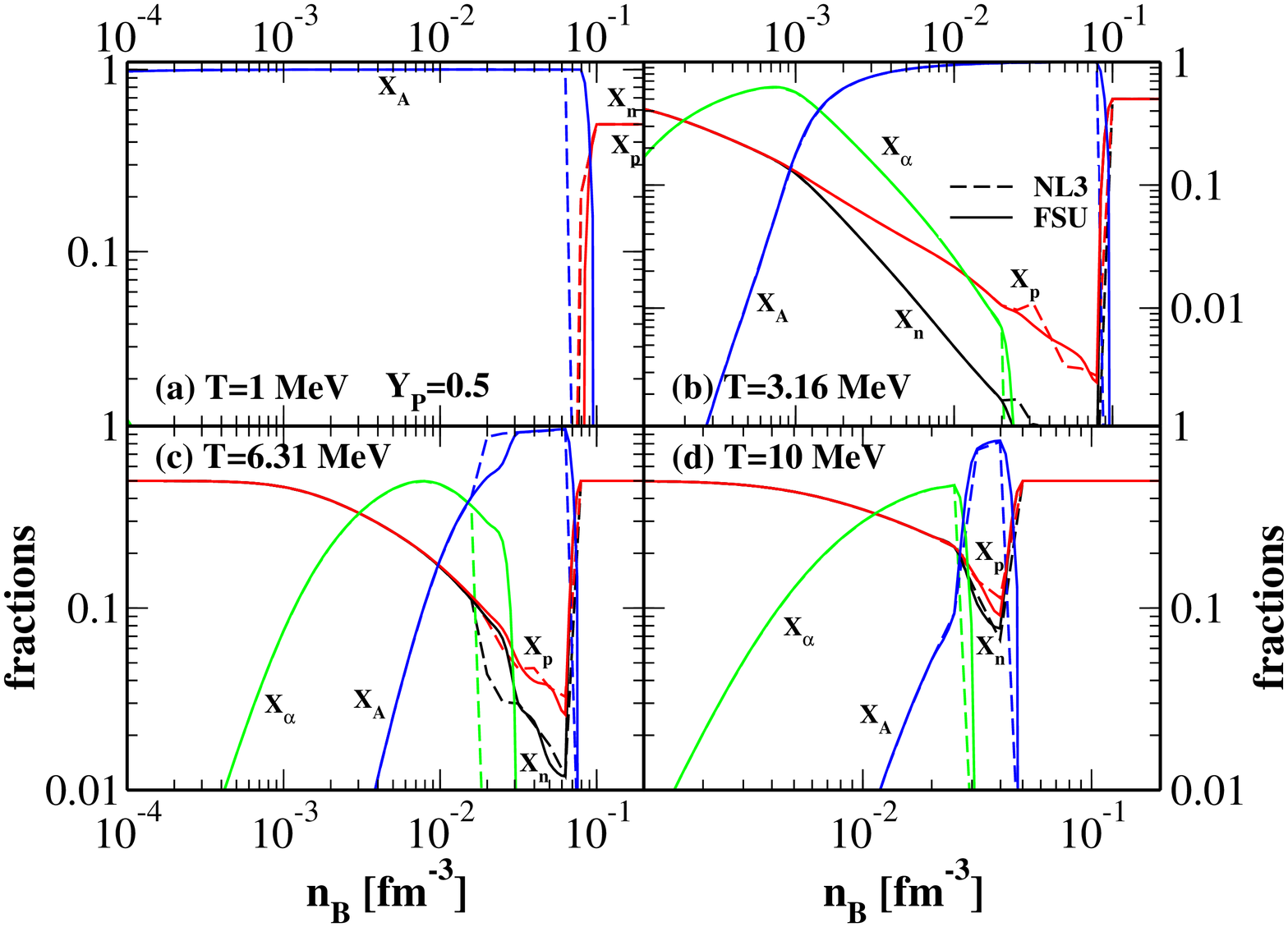}

\caption{(Color online) Mass fractions of unbound neutrons, unbound protons, $\alpha$
particles and heavy nuclei versus baryon density at four different
temperatures $T$ = 1 (a), 3.16 (b), 6.31 (c), and 10 (d) MeV, and fixed proton fraction $Y_p$ =
0.5. The solid lines are for FSU(1.7) EOS while the dashed lines are for NL3 EOS.}\label{fig:fractiony5}
\end{figure}


\begin{figure}[htbp]
 \centering
 \includegraphics[height=13cm,angle=-90]{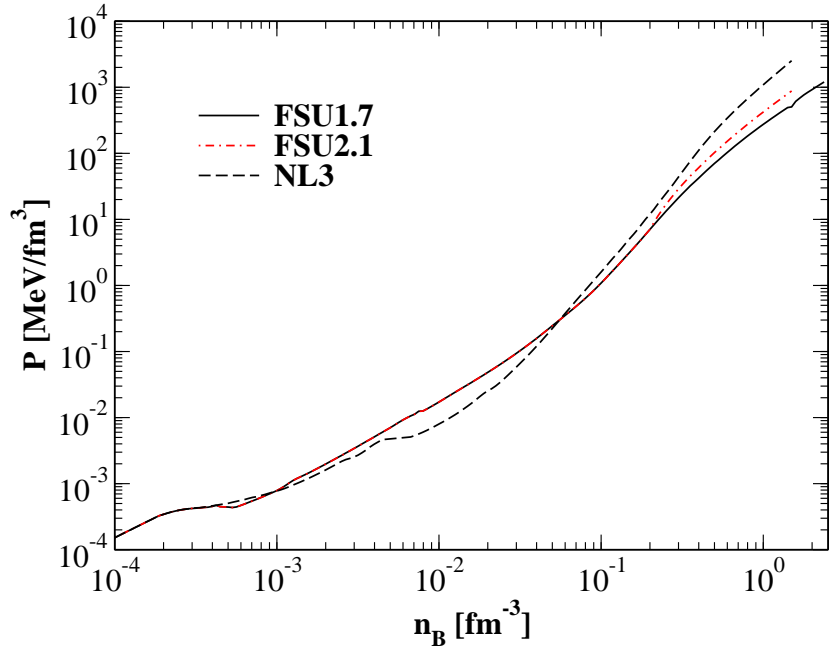}

\caption{(Color online) Pressure versus density for
zero temperature nuclear matter in beta equilibrium, for FSU1.7, FSU2.1 and NL3 EOS.
}\label{fig:T0beta}
\end{figure}

\begin{figure}[htbp]
 \centering
 \includegraphics[width=13cm,angle=0]{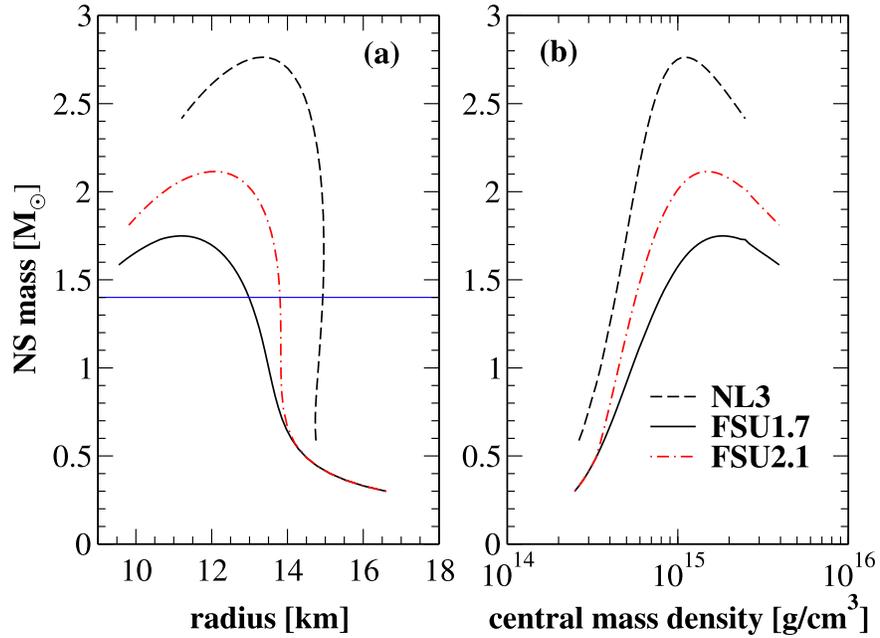}

\caption{(Color online) Mass versus radius, left (a), and mass versus central density, right
(b), for NSs using FSU1.7 (solid), FSU2.1 (dot-dashed) and NL3 (dashed) EOS as shown in
Fig.~\ref{fig:T0beta}.}
\label{fig:m-r}
\end{figure}

\begin{figure}[htbp]
 \centering
 \includegraphics[height=14cm,angle=-90]{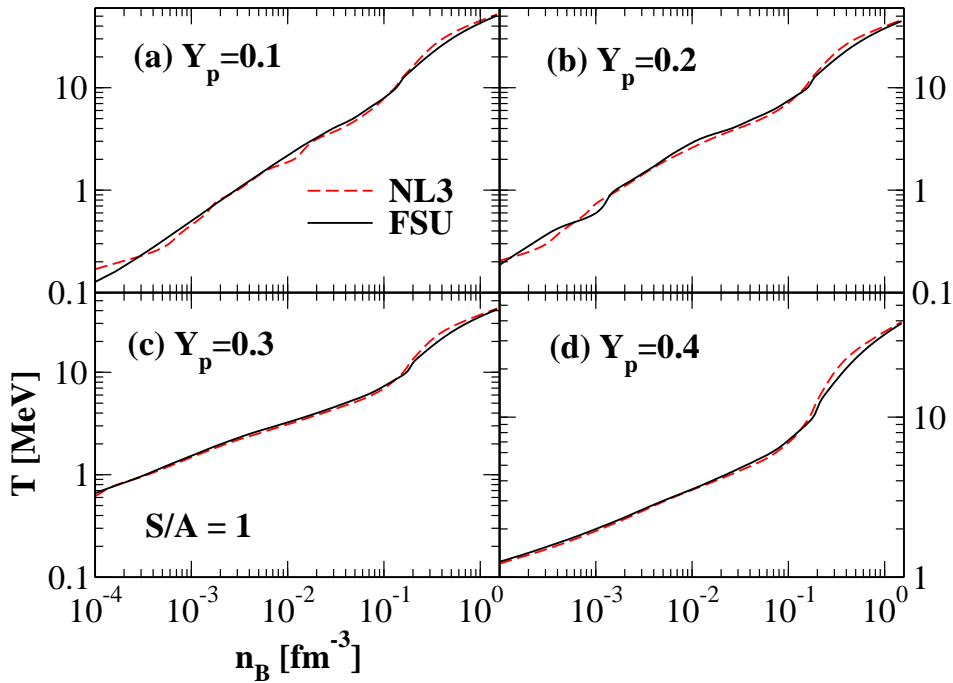}
 
\caption{(Color online) Temperature of adiabat with $S$ = 1 for different proton fractions: (a) 0.1, (b) 0.2, (c) 0.3, and (d) 0.4. Dashed (red) lines are for NL3 based EOS and solid (black) lines are for FSUGold EOS in this work.}\label{fig:s=1a}
\end{figure}

\begin{figure}[htbp]
 \centering
 \includegraphics[height=14cm,angle=-90]{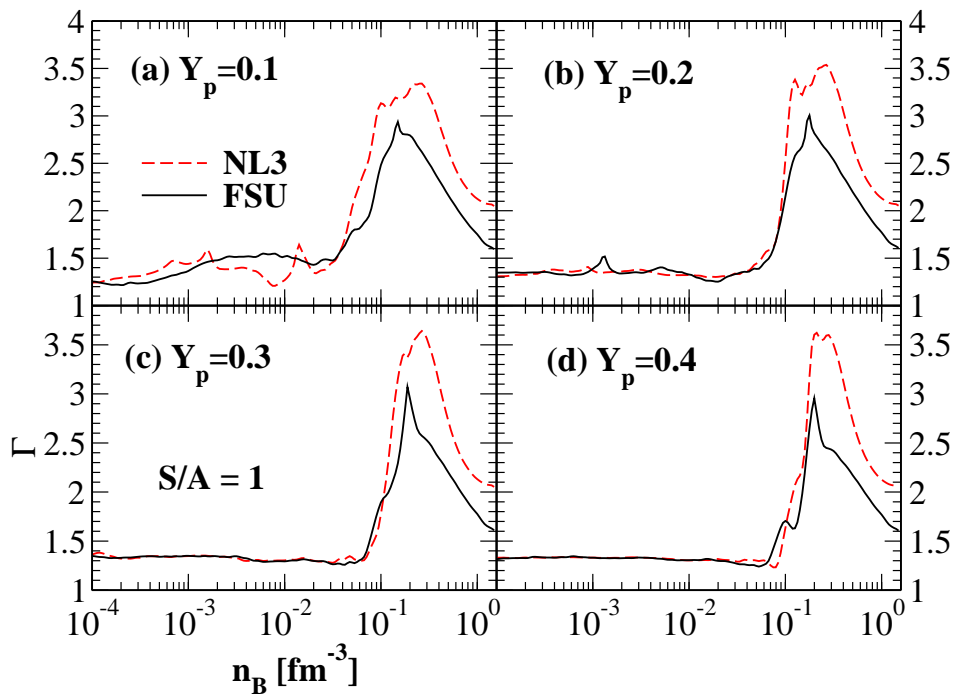}
 
\caption{(Color online) Adiabatic index $\Gamma$ along adiabat for different proton fractions $Y_p$: (a) 0.1, (b) 0.2, (c) 0.3, and (d) 0.4. Dashed (red) lines are for NL3 based EOS and solid (black) lines are for FSU1.7 EOS in this work.}\label{fig:s=1}
\end{figure}

\begin{figure}[htbp]
 \centering
 \includegraphics[scale=0.62,angle=270]{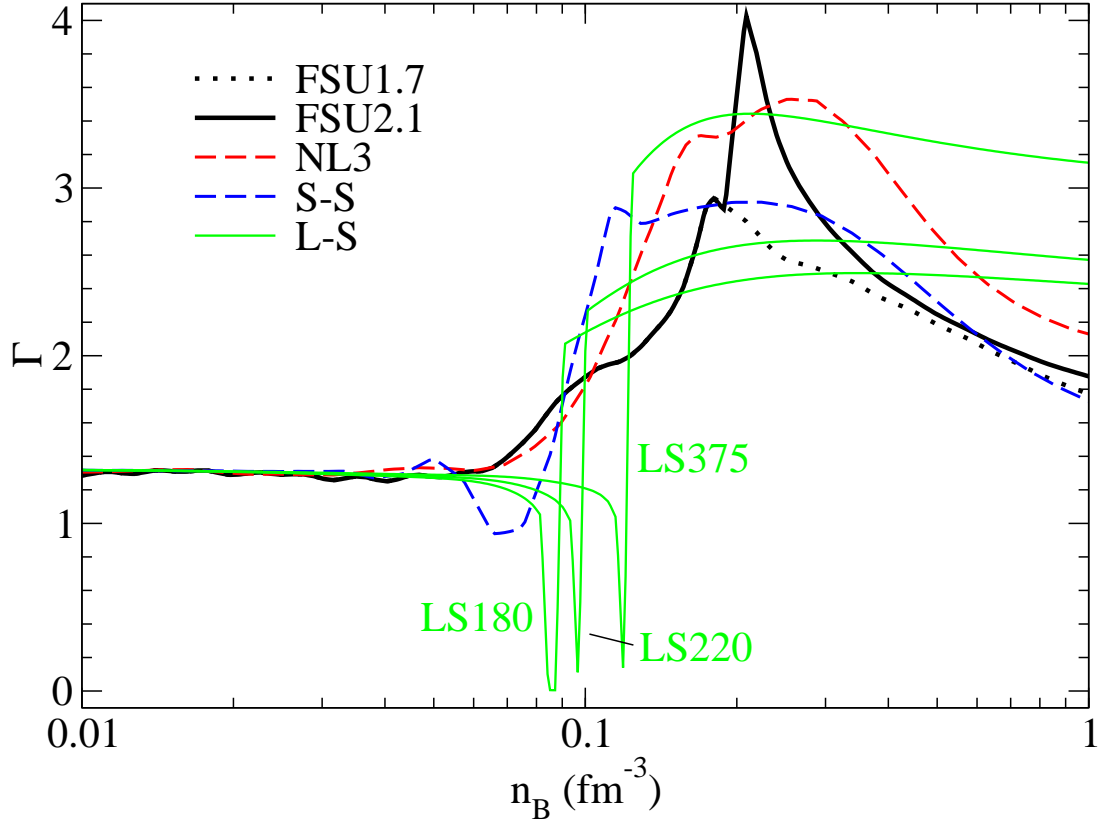}
 
\caption{(Color online) Adiabatic index $\Gamma$ for different equations of state  (EOS) assuming an initial entropy so that $T$ = 0.7 MeV at $n_B$ = 10$^{-4}$ fm$^{-3}$.  The proton fraction is $Y_p=0.3$.  The FSU1.7 and FSU2.1 EOS are from this work and are identical below $n_B=0.2$ fm$^{-3}$, while NL3 is from ref. \cite{paper3} and S-S is from ref. \cite{Shen98}.  For the Lattimer-Swesty EOS \cite{LS} three curves are shown for different versions with incompressibilities $K_0$ of 180, 220, and 375 MeV as indicated.}\label{fig:gamma}
\end{figure}


\begin{figure}[htbp]
 \centering
 \includegraphics[height=8.5cm,angle=-90]{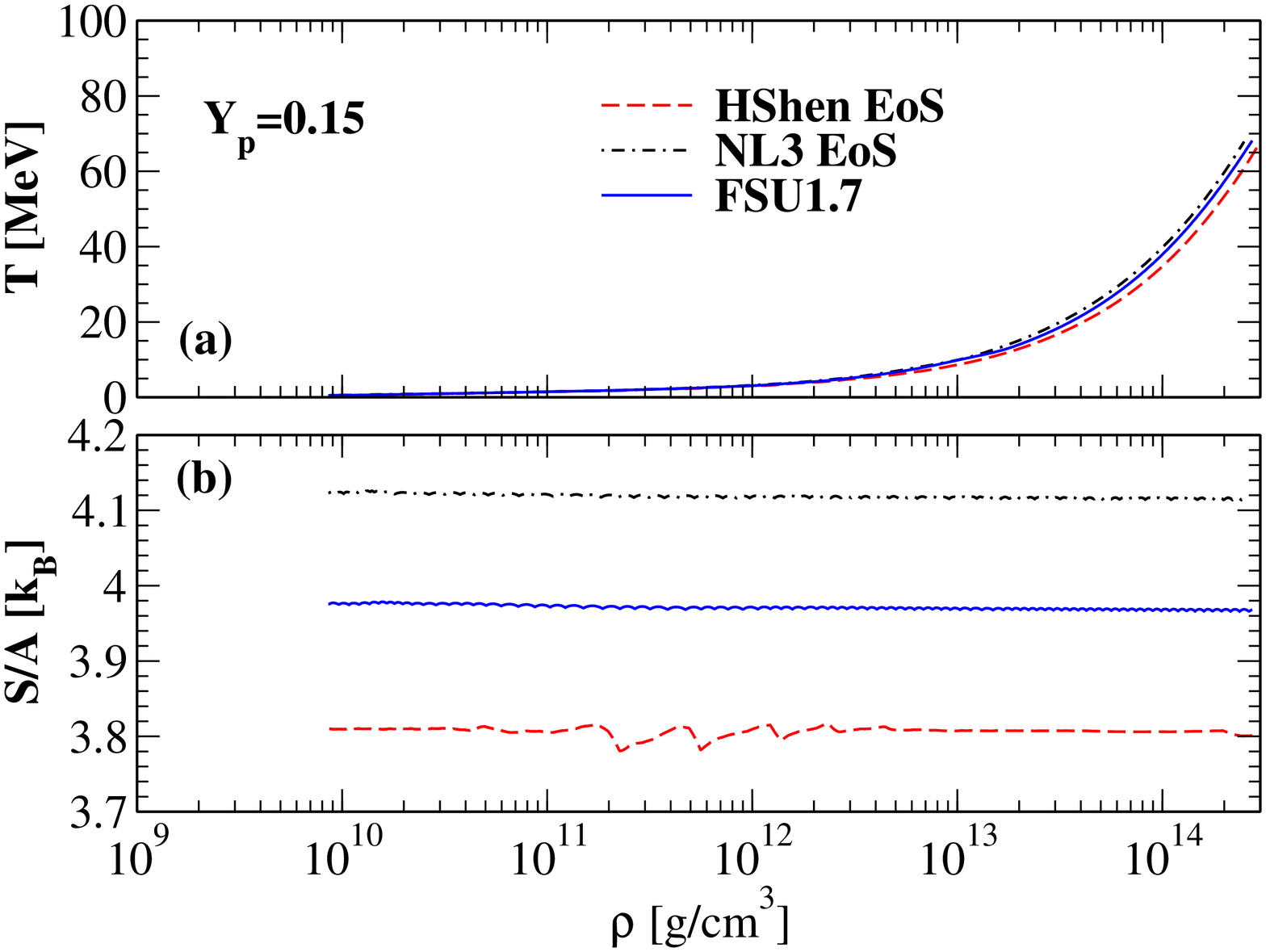}
 \includegraphics[height=8.5cm,angle=-90]{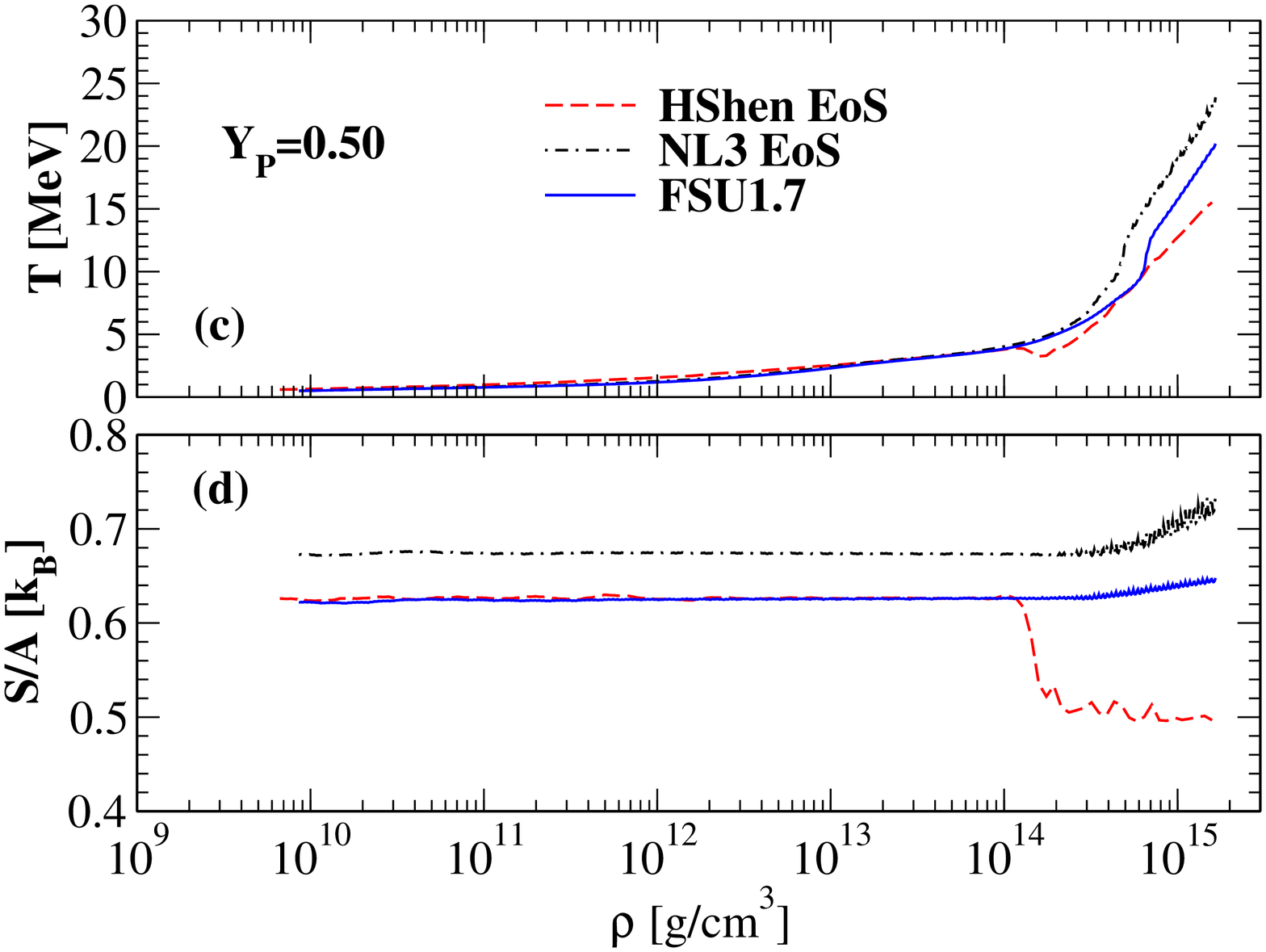}

\caption{(Color online) Left panel: Temperature (a) and entropy (b) versus density in the adiabatic compression for nuclear matter with fixed proton fraction 0.15, for HShen EOS, NL3 EOS and FSU1.7 EOS. Right panel: same as left but with proton fraction 0.5.}\label{fig:compressiony}
\end{figure}


\begin{figure}[htbp]
 \centering
 \includegraphics[height=13cm,angle=-90]{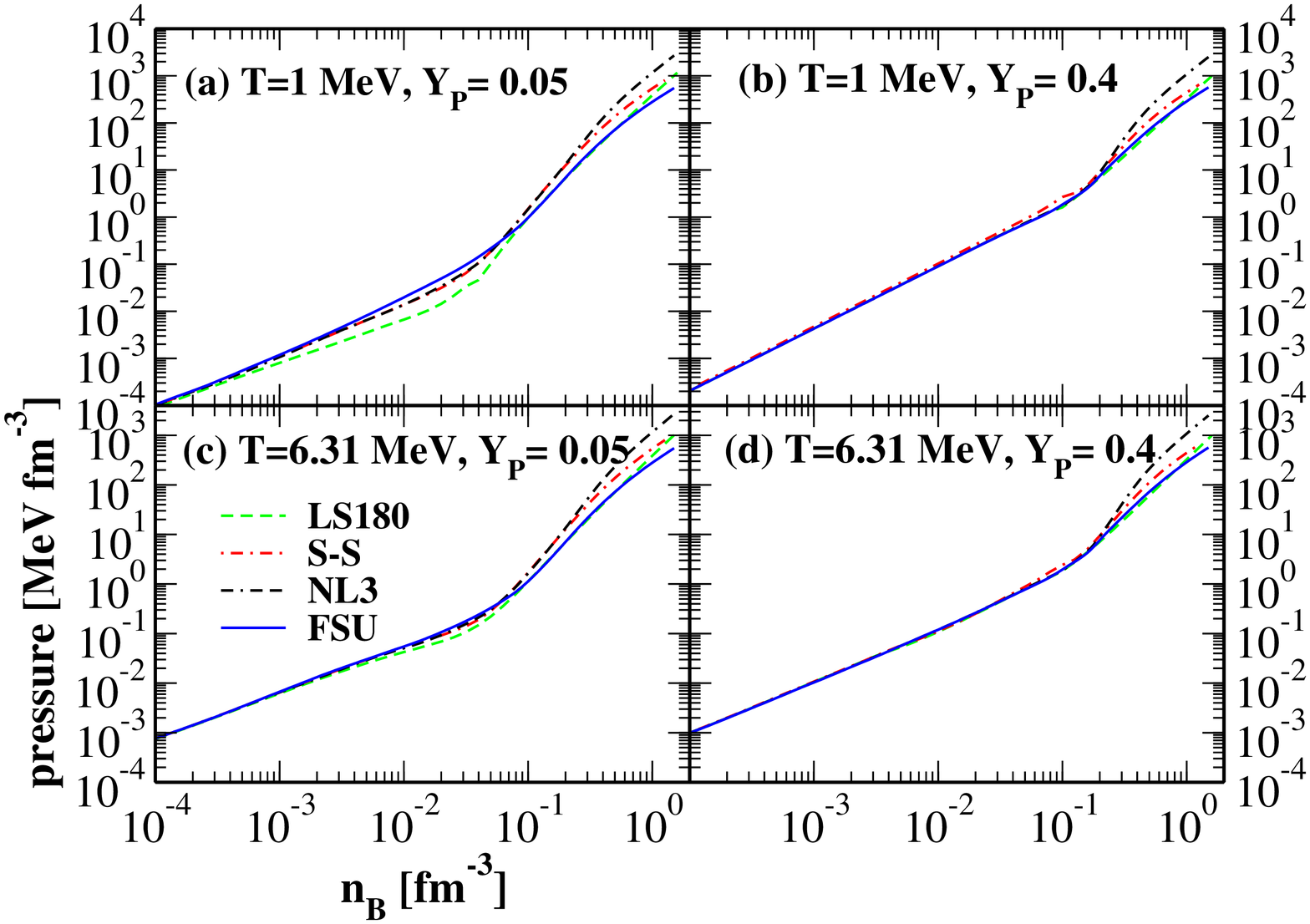}

\caption{(Color online) Pressure versus baryon density from our EOS,
Lattimer-Swesty's (L-S) and H. Shen \etal's (S-S)
EOS, with $T$ = 1 MeV,  $Y_p$ = 0.05 (a), $T$ = 1 MeV,  $Y_p$ = 0.4 (b), $T$ = 6.31 MeV,  $Y_p$ = 0.05 (c), and $T$ = 6.31 MeV,  $Y_p$ = 0.4 (d).}\label{fig:p_comp}
\end{figure}

\end{widetext}

In Fig.~\ref{fig:fractiony05}, the mass fractions of unbound neutrons,
unbound protons, $\alpha$ particles and heavy nuclei are shown versus baryon density  for a proton fraction of $Y_p$ = 0.05.  The solid lines are for the FSU EOS while the dashed lines are for NL3 EOS.  Note that for Hartree mean field calculations there is only a single nucleus associated with each Wigner-Seitz cell.   We define a nuclear level to be unbound when it has positive energy.  The upper panels (a), (b) are for $T$ = 1 and 3.16 MeV. The difference between NL3 and FSU EOS appears around 3$\times$ 10$^{-4}$ fm$^{-3}$ for $T$ = 1 MeV and 10$^{-3}$ fm$^{-3}$ for $T$ = 3.16 MeV, when the Virial gas to Hartree mean field transition occurs.  One can see that FSU and NL3 EOS differ in the mass fractions of heavy nuclei in the Hartree calculations. For $T$ = 3.16 MeV, FSU and NL3 EOS also differ in the mass fractions of protons and $\alpha$ particles, due to different transition densities. The bottom panels (c), (d) give the mass fractions of different species at higher temperatures, $T$ = 6.31 and 10 MeV. The differences between FSU and NL3 EOS are small at high temperatures.
Fig.~\ref{fig:fractiony5} shows the mass
fractions as were shown in Fig.~\ref{fig:fractiony05}, but
for matter with higher proton fraction $Y_p$ = 0.5.  Again, there are small differences between the FSU and NL3 EOS for the mass fractions, particularly for $\alpha$ particles, due to their different transition densities from Virial to Hartree.

\begin{widetext}

\begin{figure}[htbp]
 \centering
 \includegraphics[height=13cm,angle=-90]{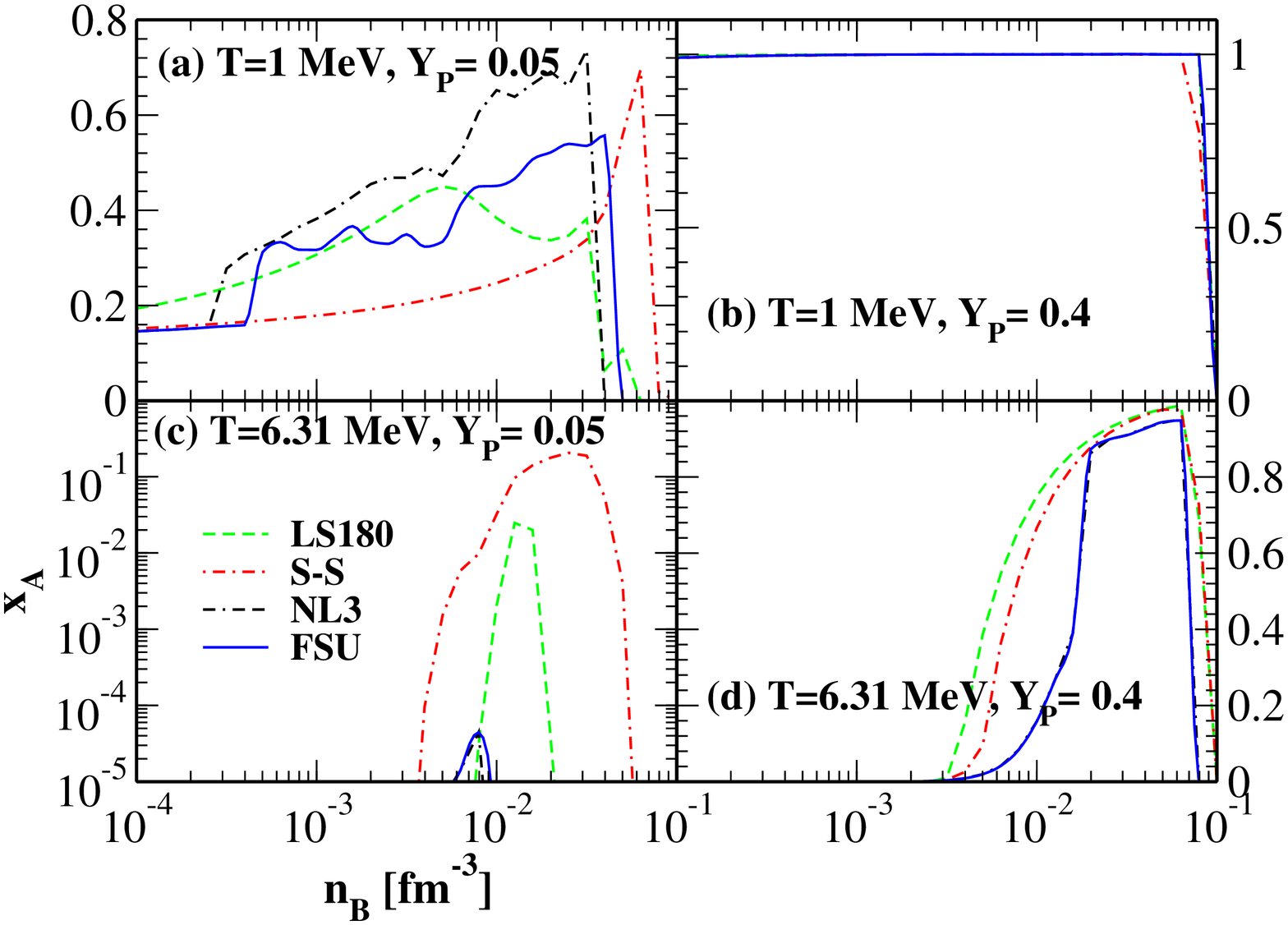}

\caption{(Color online) mass fraction of heavy nuclei versus baryon density from
our EOS, Lattimer-Swesty's and H. Shen \etal's
EOS, with $T$ = 1 MeV,  $Y_p$ = 0.05 (a), $T$ = 1 MeV,  $Y_p$ = 0.4 (b), $T$ = 6.31 MeV,  $Y_p$ = 0.05 (c), and $T$ = 6.31 MeV,  $Y_p$ = 0.4 (d).}\label{fig:xh_comp}
\end{figure}

\begin{figure}[htbp]
 \centering
 \includegraphics[height=16cm,angle=-90]{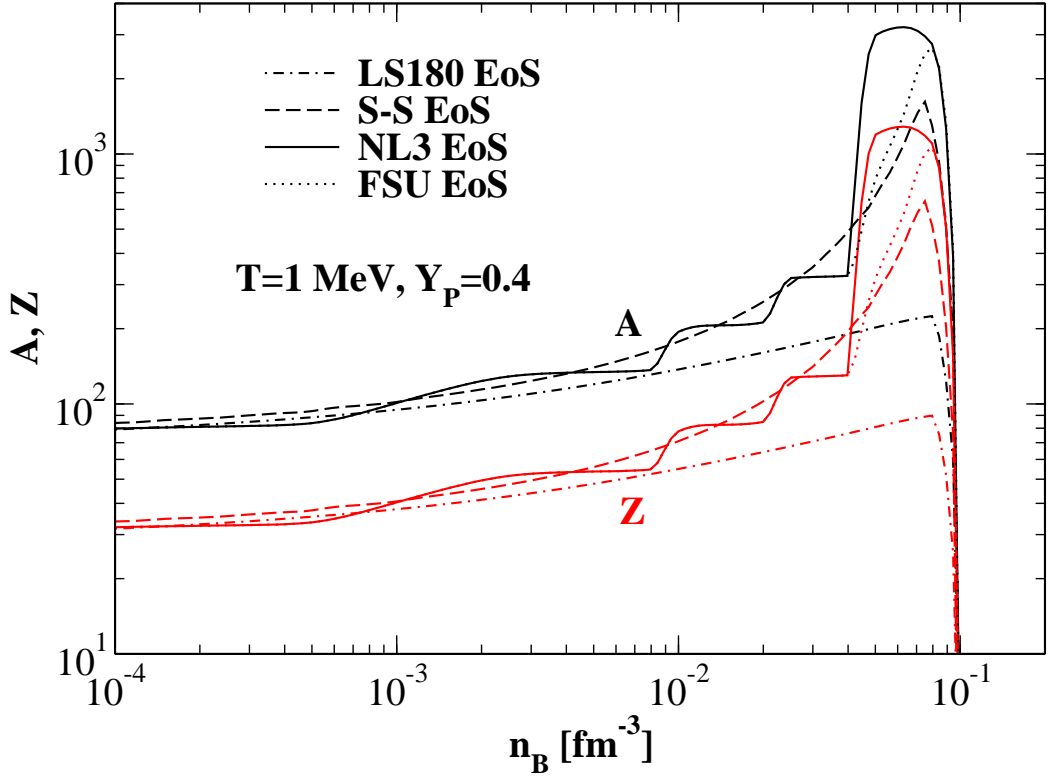}

\caption{(Color online) Average mass number $A$ (black upper curves) and atomic number $Z$ (red lower curves) of heavy nuclei for the FSU EOS (double dashed-dot), NL3 EOS (solid) , Lattimer-Swesty's (dot-dashed) and H. Shen \etal's EOS (dashed) for $T$ = 1 MeV,  $Y_p$ = 0.4.}\label{fig:AZ}
\end{figure}

\begin{figure}[htbp]
 \centering
 \includegraphics[height=13cm,angle=-90]{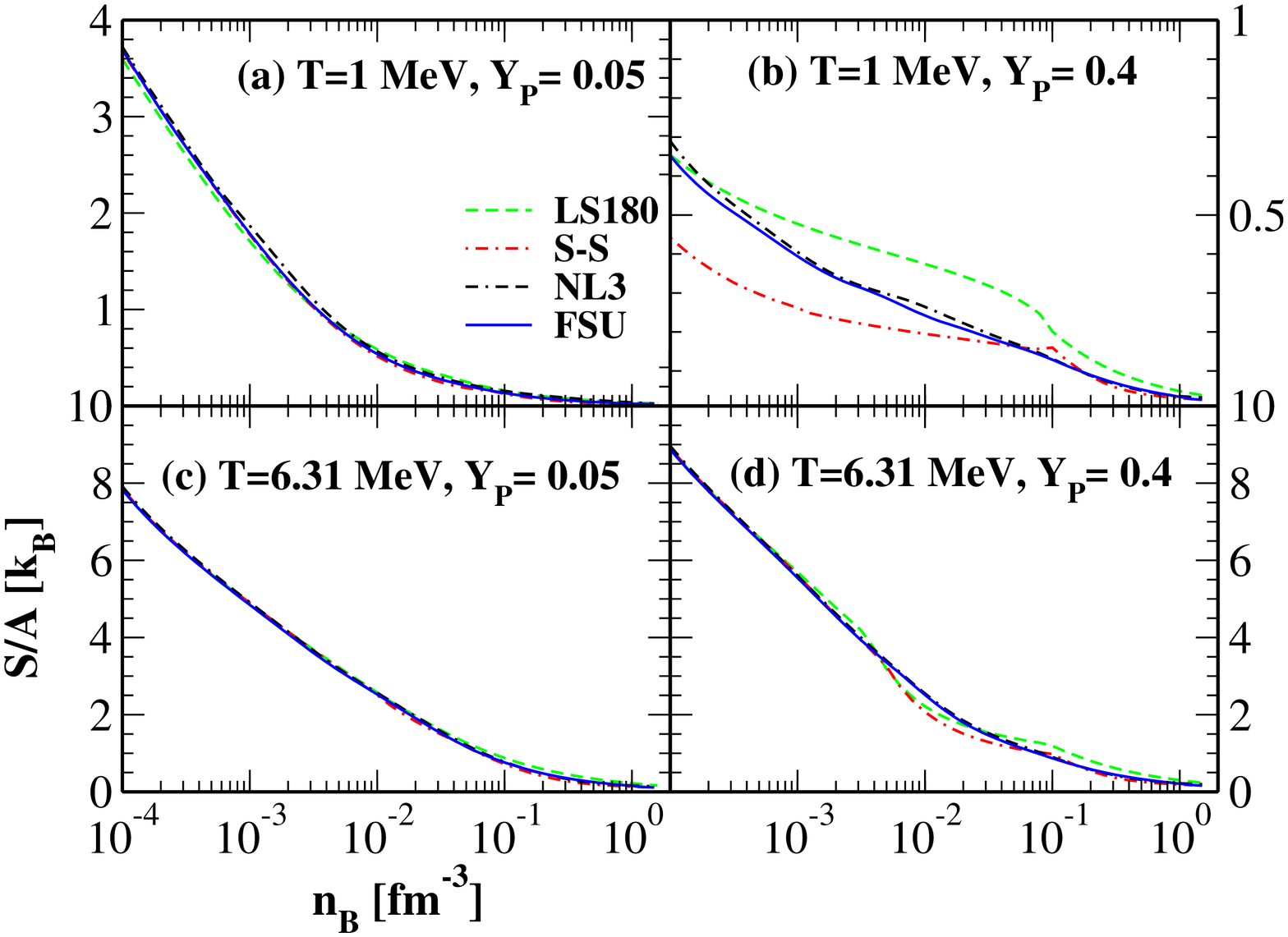}

\caption{(Color online) Entropy per baryon versus baryon density from our EOS,
Lattimer-Swesty's and H. Shen \etal's EOS, with $T$ = 1 MeV,  $Y_p$ = 0.05 (a), $T$ = 1 MeV,  $Y_p$ = 0.4 (b), $T$ = 6.31 MeV,  $Y_p$ = 0.05 (c), and $T$ = 6.31 MeV,  $Y_p$ = 0.4 (d).}\label{fig:s_comp}
\end{figure}

\end{widetext}

\subsection{Neutron Star Structure}

Here we present the zero temperature EOS of nuclear matter in beta equilibrium and the resulting NS mass radius relation. The results for the FSU1.7, FSU2.1, and NL3 EOS are compared.  Fig.~\ref{fig:T0beta} shows the pressure versus density for zero temperature nuclear matter in beta equilibrium. The FSU1.7 and FSU2.1 EOS are stiffer (higher pressure) than the NL3 EOS for intermediate densities above 10$^{-3}$ fm$^{-3}$, and then softer than NL3 when the density is above 5$\times 10^{-2}$ fm$^{-3}$.  FSU2.1 becomes stiffer than FSU1.7 when the density is above 0.2 fm$^{-3}$. The stiffness of the EOS at high density determines the maximum NS mass. Therefore FSU1.7, FSU2.1, and NL3 predict maximum NS masses of 1.7, 2.1, and 2.8 $M_\odot$ respectively, as shown in the left panel (a) of Fig.~\ref{fig:m-r}.  The FSU1.7, FSU2.1, and NL3 EOS predict radii of roughly 13 km, 14 km, and 15 km, respectively, for a 1.4 $M_\odot$ NS. The right panel (b) of Fig.~\ref{fig:m-r} shows the NS mass vs central density.  For a NS with given mass, the central density of the NS is inversely related to the stiffness of the EOS at high density.

\subsection{Adiabatic index}

In this subsection we discuss the adiabatic index $\Gamma$ of nuclear matter, at constant entropy, for several different EOS.  Figure~\ref{fig:s=1a} shows the temperature of the adiabat, with entropy $S/A$ = 1, for nuclear matter with proton fractions of $Y_p$ = 0.1, 0.2, 0.3, and 0.4.  Note that baryon,
electron, positron, and photon contributions to $S$ are included.  The solid lines are for the FSU1.7 EOS while the dashed lines are for the NL3 EOS. Above saturation density FSU1.7 gives a slightly higher temperature than NL3. Since the modification term, Eq.~(\ref{dp}), in FSU2.1 does not depend on temperature, there is no difference between FSU1.7 and FSU2.1 in the adiabats.

The adiabatic index $\Gamma$, 
\beq\label{adia} \Gamma\ =\ \left(
\frac{\partial \mathrm{ln}P}{\partial \mathrm{ln}n}\right)_s, 
\eeq
describes the stiffness of the EOS at constant entropy.  In Fig.~\ref{fig:s=1}, the adiabatic index is shown versus density for nuclear matter with constant entropy $S/A$ = 1 and  $Y_p$ = 0.1, 0.2, 0.3, and 0.4.  At subnuclear density, $\Gamma$ has only small fluctuations with density. It then rises rapidly at the transition from nonuniform to uniform matter.  This characterizes the stiffening of the EOS due to the large nuclear incompressibility.  Beyond saturation density, NL3 has a larger value of $\Gamma$ than FSU1.7. For FSUGold EOS, we incorporated more data for Virial EOS at low temperatures, which leads to small variations in the numerical evaluation of pressure and entropy at low density. This induces small differences between NL3 and FSU EOS in the adiabatic index and temperature at low densities.

In Fig.~\ref{fig:gamma} we present the adiabatic index of various EOS assuming an initial entropy so that $T$ = 0.7 MeV at a density $n_B$ = 10$^{-4}$ fm$^{-3}$ for $Y_p=0.3$.    All EOS show a rapid rise of $\Gamma$ beyond the transition to uniform matter $\sim$ 0.1 fm$^{-3}$, due to the incompressibility of nuclear matter.  There are different versions of the Lattimer-Swesty EOS depending on the assumed value for the incompressibility of nuclear matter.  We refer to LS180, LS220, and LS375 for the Lattimer-Swesty EOS with $K_0=180, 220$, and 375 MeV, respectively.
The LS180, LS220, LS375 and H. Shen EOS have a significant dip in $\Gamma$ at densities just below the transition to uniform matter. In contrast the NL3, FSU1.7, and FSU2.1, EOS, do not show an obvious dip.

The Lattimer-Swesty EOS is based on a simple liquid drop model and assumes a first order phase transition between a low density vapor phase and a high density liquid phase.  During this transition the pressure is independent of density so $\Gamma$ is small.  Indeed a first order phase transition may be appropriate for small systems such as heavy ion collisions.  For small systems, Coulomb effects are relatively mild so that one can form a single uniform liquid phase.   However, for large astrophysical systems, Coulomb interactions play a crucial role and prevent the formation of a uniform liquid phase.  Instead, one must form a nonuniform phase with average proton density equal to the electron density.  For this nonuniform phase, our relativistic mean field calculations give an adiabatic index that increases with density and does not show a decrease.    Therefore, we believe that the decrease in $\Gamma$ for the LS EOS may be an artifact of the approximations that they use and their assumption of a first order phase transition.  The behavior of $\Gamma$ for the H. Shen EOS may be related to their Thomas-Fermi approximation.  For example, it is not clear if they considered a full range of possible shapes for nonuniform matter such as the shell states \cite{paper0} that could be present in this density range.  We emphasize that the H. Shen EOS involves variational calculations and Thomas Fermi approximations to the relativistic mean field formalism that we calculate directly.  

The FSU1.7 and FSU2.1 EOS have a larger $\Gamma$ compared to NL3 in the density range $n_B=0.07$ to 0.1 fm$^{-3}$.  This is probably related to differences in the density dependence of the symmetry energy, see $L$ in Table \ref{Table3}.  Since $L$ is smaller for the FSU EOS, it actually has a {\it larger} symmetry energy and pressure at low densities than NL3.  This could lead to the larger $\Gamma$ at low densities.  Nuclear measurements can constrain $L$ and reduce the model uncertainty in $\Gamma$ at low densities.  For example, $L$ can be constrained by measuring the neutron radius of $^{208}$Pb with parity violating electron scattering \cite{PREX1,PREX2} or with measurements of isospin diffusion in heavy ion collisions \cite{isodiffusion}.  The possible impact of the density dependence of the symmetry energy on the EOS should be studied further.        

The rising of $\Gamma$ for FSU2.1 at $n_B$=0.2 fm$^{-3}$, due to the modification term to the pressure (and energy) beyond this density, indicates stiffening of EOS at high density. In a more consistent way to increase the pressure at high densities by refitting the effective interaction as discussed in Sec. IIB, $\Gamma$ may rise more gently with density, which we will explore in future work. However the rise of $\Gamma$ in FSU2.1 does not violate any physical principle we are aware of, especially the causality.  Moreover, the peak at 0.2 fm$^{-3}$ becomes smaller and the small kink disappears, if the EOS table has a lower resolution, for example 20 points per decade in temperature and density. 

\subsection{Adiabatic Compression Tests}
\label{sec.test.results}

It is important that an EOS table is thermodynamically consistent.  Otherwise small errors in the table can lead to artificial increases in entropy during an astrophysical simulation.  Here we present adiabatic compression test results for FSU1.7 and FSU2.1 EOS. The details of this test have been discussed in a previous paper \cite{paper3}.  The entropy should be conserved in this adiabatic compression independent of the density.  In Fig.~\ref{fig:compressiony}, the temperature and entropy versus density during  adiabatic compression are shown for nuclear matter with fixed proton fraction $Y_p =$ 0.15 (left) and 0.50 (right).  The initial density is $5.16738\times 10^{-6}$ fm$^{-3}$ and initial $T$ = 0.5 MeV. The FSU and NL3 EOS conserve entropy within 1\%, except for a slight rise in $S/A$ at extremely high density when $Y_p$=0.5. For a comparison test results for the H. Shen EOS is also shown (as red dashed curve) for similar initial conditions. Note that it is important to use an accurate interpolation scheme with the EOS table to ensure that the first law is satisfied so that entropy is conserved. The adiabatic compression test result for the H. Shen EOS was obtained using the routine developed in Ref.~\cite{Ott09}. In typical adiabatic compression tests, one has to interpolate the existing tables. For nuclear matter at high densities with high $Y_p$, the thermodynamic properties change fast as density rises. The NL3, FSU, HShen EOS used for this plot have 40 points per decade in density and temperature axes. For a finer table with 80 points per decade, we find the entropy is conserved better at high densities(within 1 \% deviation). However, this would increase the volume of table by a factor of four. Since this region is most likely not relevant in typical supernova simulations, we decided to generate the smaller 40 points per decade table.

\subsection{Comparisons with existing EOS}

It is useful to compare the thermodynamics of the new FSU EOS with the NL3 EOS, the Lattimer-Swesty EOS (L-S), that uses a simple liquid drop model, and the H. Shen \etal\ EOS (S-S), that uses the Thomas Fermi approximation and variational calculations with approximating shapes for nuclei. The L-S EOS quoted in this subsection corresponds to the one with incompressibility coefficient $K_0$=180 MeV, {\it i.e.} LS180. The comparisons for pressure, entropy, and heavy nuclei fractions between NL3, LS180, and S-S EOS have been discussed in a previous paper \cite{paper3}. In this subsection, we will focus on the differences between the FSU1.7 and the other EOS. The only difference between FSU1.7 and FSU2.1 EOS is the pressure and energy at densities above 0.2 fm$^{-3}$, as explained in Sec.~\ref{modifiedFSU}.

In Fig.~\ref{fig:p_comp}, the pressure for matter at
$T$ = 1, or 6.31 MeV and $Y_p$ = 0.05 or 0.4 is shown for the FSU
EOS, NL3 EOS, LS180 and S-S EOS. The pressure includes contributions from electrons, positrons, and photons. For neutron rich matter at subnuclear density as in the left upper panel (a) for $T$=1 Mev and $Y_p$=0.05, the FSU EOS gives slightly larger pressure than all the other EOS. As the temperature and/or proton fraction grows, this difference becomes less obvious. On the other hand, for nuclear matter above saturation density, the FSU EOS gives the smallest pressure among all of these EOS.

Fig.~\ref{fig:xh_comp} compares the mass fraction of heavy nuclei from the
FSU EOS, NL3 EOS, LS180 EOS and S-S EOS, for matter at $T$ = 1 or 6.31 MeV,
and $Y_p$ = 0.05 or 0.4. The mass fraction of heavy nuclei can be very different among the various EOS and this causes the differences in entropy. However, the FSU and NL3 EOS only differ in the mass fraction of heavy nuclei for neutron rich matter at subnuclear density and low temperature as shown in the left upper panel for $T$=1 Mev and $Y_p$=0.05. This is also discussed in Fig.~\ref{fig:fractiony05}. Note possible reasons for the differences in the mass fractions between NL3, LS180 and S-S EOS have been discussed in the previous paper \cite{paper3}.  The sharp rise in the mass fraction for FSU and NL3 EOS with T = 6.31 MeV and $Y_p$ = 0.4, also T = 1 MeV and $Y_p$ = 0.05, is due to transition from Virial to RMF results. For RMF, we solved the nuclear energy levels in non-uniform nuclei exactly and denoted nucleon as “free” when its energy is positive. This is different from Virial EOS and it is possible the composition is not smooth everywhere.
 The average $A$ and $Z$ of heavy nuclei for $T$ = 1 MeV and $Y_p$ = 0.4 is shown in Fig.~\ref{fig:AZ}.  This figure shows the effects of shell structure, that is included in FSU and NL3 EOS but is neglected in both LS180 and S-S EOS. As a result, $A$ and $Z$ for our EOS have a series of steps while $A$ and $Z$ for LS180 and S-S EOS increase smoothly with density. Note the NL3 EOS gives larger $A$ and $Z$ in the Hartree non-uniform region compared to the FSU EOS. NL3 also shows somewhat of a plateau in $A$ and $Z$, where nuclei are in the shell shape \cite{paper0}, while the FSU EOS does not give shell shape nuclei for this value of $Y_p$ (it does for most values of $Y_p$). More detailed discussion about shell shape nuclei can be found in Ref.~\cite{paper0}.

Fig.~\ref{fig:s_comp} compares the entropy from
FSU EOS, NL3 EOS, LS180 EOS and S-S EOS, for nuclear matter at $T$ = 1 or 6.31 MeV,
and $Y_p$ = 0.05 or 0.4. Note that heavy
nuclei have low entropy at low temperatures so the difference
in entropy between the FSU and NL3 EOS is small. While at higher temperatures, the entropy is dominated by unbound nucleons and photons and the two EOS agree very well.

\section{Summary}

In this paper we present a second equation of state (EOS) of nuclear
matter for a wide range of temperatures, densities, and proton
fractions for use in supernovae, neutron star - neutron star mergers, neutron star - black hole mergers, and black hole formation
simulations.  We use a density dependent relativistic mean field (RMF) model, FSUGold, for nuclear
matter at intermediate and high density with a spherical
Wigner-Seitz approximation for nonuniform matter, which
incorporates nuclear shell effects \cite{paper1}. For nuclear matter at
low density \cite{paper2}, we use a Virial expansion for a non-ideal gas consisting
of neutrons, protons, $\alpha$ particles and thousands of heavy
nuclei from the finite range droplet model (FRDM) mass table \cite{FRDM}. The difference from the first EOS is that here we use the RMF effective interaction FSUGold whereas our first EOS was based on the NL3 effective interaction. The Virial gas part is common and has been discussed in a previous paper \cite{paper2}.

We tabulate the resulting EOS at over 100,000 grid
points in the temperature range $T$ = 0 to 80 MeV, the density
range $n_B$ = 10$^{-8}$ to 1.6 fm$^{-3}$, and the proton
fraction range $Y_p$ = 0 to 0.56.  We present differences between our FSUGold based EOS and our NL3 based EOS along with some existing EOS, for the thermodynamic properties, composition, and neutron star structure. In particular, we focus on the differences between the FSU and NL3 EOS. The FSUGold EOS is considerably softer than NL3 EOS in the high density, but stiffer at subnuclear density, as shown in Fig.~\ref{fig:p_comp}. The FSU EOS also predicts different mass fractions and average $A$ and $Z$ of heavy nuclei from NL3 EOS, as shown in Figs.~\ref{fig:fractiony05},    \ref{fig:fractiony5}, and \ref{fig:AZ}.

As for the NL3 EOS, we use a hybrid interpolation scheme to generate a full table for the FSU EOS on a fine grid that is thermodynamically consistent.  This ensures that the first law of thermodynamics is satisfied and that entropy is conserved during adiabatic compression.

Moreover, the original FSUGold EOS has a maximum neutron star mass of 1.7 $M_\odot$. We introduce a modification in the pressure at high density, which accommodates a 2.1 $M_\odot$ neutron star as shown in Fig.~\ref{fig:m-r}. The EOS based on the original FSUGold is therefore called FSU1.7 while the modified EOS is called FSU2.1, according to their different maximum NS masses. Note in the final EOS tables the upper limit in densty is 10$^{0.175}$ and 10$^{0.375}$ fm$^{-3}$ for FSU2.1 and FSU1.7, respectively.

Finally, the EOS tables for FSU1.7 and FSU2.1 are available for download as described in Appendix \ref{app}. Our goal in constructing EOS tables is two fold.  First, we intend to calculate EOS tables with different pressures.  This will allow one to correlate features of astrophysical simulations with EOS properties.  In this paper we calculate a new EOS that has a lower pressure, at high densities, than our previous NL3 EOS. Second, we aim to provide detailed composition information in future EOS tables.  This can be important for neutrino interactions.  Historically, most EOS tables, used in astrophysical simulations, have only provided mass fractions for neutrons, protons, $\alpha$s, and a single average heavy nucleus.  However, electron capture on a range of heavy nuclei may be important for the proton fraction $Y_p$ during the infall phase of a supernova \cite{electroncapture}.  In addition, mass 3 and other light nuclei can be important for antineutrino opacities \cite{mass3virial,mass3}. Deuterons have been included in current Virial EOS via the second Virial coefficient between neutrons and protons \cite{paper2}, and we plan to include mass 3 and other light nuclei in future work. Note that the EOS table described in this paper only provides a single average heavy nucleus.  This is done to be compatible with existing simulations and neutrino opacity codes.  However we will make available additional composition information in future work.

\section{Acknowledgement}

We thank Lorenz H$\ddot{\mathrm{u}}$depohl, Thomas Janka, Andreas Marek, and Christian Ott for important help running astrophysical simulations to debug our equation of state, and Scott Teige for help on shell script for running job on Teragrid supercomputer cluster Ranger. This work was supported in part by DOE grant DE-FG02-87ER40365, and Teragrid grant PHY100015 for computing time. The work of GS was also supported in part by a grant from the DOE under contract DE-AC52-06NA25396 and the DOE topical collaboration to study "Neutrinos and nucleosynthesis in hot and dense matter".

\section{\label{app}Appendix: format of EOS tables}

Here we describe where the tables
FSU1.7EOSb1.01.dat FSU1.7EOS1.01.dat, FSU2.1EOS1.01.dat and FSU2.1EOS1.01.dat can be downloaded. The entries in these tables are the same as those in NL3 EOS tables and explained in a previous paper \cite{paper3}. One should download the gzip compressed files (that are about 100 MB each) and use gunzip to decompress them.  The grid structures of these tables are indicated in Table \ref{tab:phasespace2} and contain approximately 517 MB of data each.  The tables, a sample FORTRAN computer program, and a readme file are available for download both at our website 
\begin{widetext}
\underline{\url{http://cecelia.physics.indiana.edu/gang_shen_eos/}} 
\smallskip
\end{widetext}
and, from the  Electronic Physics Auxiliary Publication Service - EPAPS web site \cite{EPAPS}.   Please check our web site for any updated information regarding these EOS tables. The entries in the table are the same as that for the NL3 EOS and have been discussed in the appendix of Ref.~\cite{paper3}.

\end{document}